\documentclass[aps,prstab,twocolumn,groupedaddress,floatfix,showpacs]{revtex4}
\usepackage{amsmath,amsfonts,amssymb,graphics,graphicx,epsfig,color}
\newcommand{\refeq}[1]{~(\ref{#1})}
\newcommand{\myref}[1]{~\ref{#1}}

\newcommand{\eeq}{\end{equation}}

\begin{document}

\title{{\large\bf A stochastic-hydrodynamic model
of halo formation in charged particle beams}\\
\vspace{8pt}{\small\rm (revision day: February 15, 2003)}}

\author{Nicola Cufaro Petroni}
\email[]{cufaro@ba.infn.it} \affiliation{Dipartimento Interateneo
di Matematica -- Universit\`a e Politecnico di Bari;\\
T.I.R.E.S. (Innovative Technologies for Signal Detection and Processing)
-- Universit\`a di Bari; \\
Istituto Nazionale di Fisica Nucleare -- Sezione
di Bari;\\
Istituto Nazionale per la Fisica della Materia --
Unit\`a di Bari;\\
Via G. Amendola 173, 70126 Bari, Italy.}
\author{Salvatore De Martino}
\email[]{demartino@sa.infn.it}
\author{Silvio De Siena}
\email[]{desiena@sa.infn.it}
\author{Fabrizio Illuminati}
\email[]{illuminati@sa.infn.it} \affiliation{Dipartimento di
Fisica ``E. R. Caianiello'' --
Universit\`a di Salerno;\\
Istituto Nazionale per la Fisica della Materia --
Unit\`a di Salerno;\\
Istituto Nazionale di Fisica Nucleare -- Sezione
di Napoli (Gruppo collegato di Salerno);\\
Via S. Allende, I--84081 Baronissi (SA), Italy.}

\begin{abstract}
The formation of the beam halo in charged particle accelerators
is studied in the framework of a stochastic-hydrodynamic
model for the collective motion of the particle beam.
In such a stochastic-hydrodynamic theory the density and
the phase of the charged beam obey a set of coupled nonlinear
hydrodynamic equations with explicit time-reversal invariance.
This leads to a linearized theory that describes the collective
dynamics of the beam in terms of a classical Schr\"odinger
equation.
Taking into account space-charge effects, we derive
a set of coupled nonlinear hydrodynamic equations.
These equations define a collective dynamics of self-interacting
systems much in the same spirit as in the Gross-Pitaevskii
and Landau-Ginzburg theories of the collective dynamics for
interacting quantum many-body systems.
Self-consistent solutions of the dynamical equations
lead to quasi-stationary
beam configurations with enhanced transverse dispersion
and transverse emittance growth. In the limit of a
frozen space-charge core it is then possible to
determine and study the properties of stationary, stable
core-plus-halo beam distributions.
In this scheme the possible reproduction of the halo
after its elimination is a consequence of the stationarity of the
transverse distribution which plays the role of an attractor for
every other distribution.
\end{abstract}


\maketitle \vspace{1cm}

\section{Introduction}\label{introduction}

High intensity beams of charged particles, in particular in
linacs, have been proposed in recent years for a wide variety of
accelerator--related applications: drivers for sources of neutron
spallation; production of tritium; transmutation of radioactive
wastes to species with shorter lifetimes; heavy ion drivers for
fusion-based production of thermonuclear energy; production of
radioactive isotopes for medical use, and so on. In all these
cases it is very important to keep at a low level the beam losses
to the wall of the beam pipe, since even small fractional losses
in a high--current machine can cause exceedingly high levels of
radioactivation. It is now widely believed that one of the
relevant mechanisms for these losses is the formation of a low
intensity halo relatively far from the core of the beam. These
halos have been either directly observed \cite{koziol} or inferred
from experiments \cite{reiser}, and have also been predicted from
extensive numerical simulations
\cite{gluck94,gluck95,gluck96,okamoto97,gluck98,wangler98,gluck99,ikegami99,quiang00}.
Often in these studies the particle--core model has been used to
understand the halo formation and its extent. That
notwithstanding, it is widely believed that for the next
generation of high intensity machines it is necessary to obtain a
more quantitative understanding not only of the physics of the
halo, but also of the beam transverse distribution in general
\cite{hofmann00,hofmann01,hofmann02}. In fact ``because there is
not a consensus about its definition, halo remains an imprecise
term'' \cite{wangler} so that several proposals have been put
forward for its description.

Charged particle beams are usually described in terms of
classical deterministic dynamical systems.
The standard model is that of a
collisionless plasma where the corresponding dynamics is embodied
in a suitable phase space (see for example \cite{landau}).
Generally speaking, the beams are described in a comoving reference
system so that we can confine ourselves to a non relativistic
setting. In this framework, the formation of the halo has been
studied mainly by
means of the particle-core model initially proposed by Gluckstern,
and the simulations show that the dynamical instabilities due to a
parametric resonance can allow ions to escape from the
core
\cite{gluck94,gluck95,gluck96,okamoto97,gluck98,wangler98,gluck99,ikegami99,quiang00}.

In the present paper we propose and develop a different approach.
We introduce a model for the formation of the halo based on the
idea that the trajectories of the charged particles are sample
paths of a stochastic process, rather than the usual deterministic
(differentiable) trajectories. A random collisional plasma
described in terms of configurational stochastic processes seems a
realistic model for the dynamics taking place in particle
accelerators. Moreover, it can explain the rare escape of
particles from a quasi-stable beam core, since it takes into
account, statistically, the complicated inter-particle
interactions that cannot be described in detail. Of course, the
idea of a stochastic approach is hardly new
\cite{landau,ruggiero,struckmeier}, but there are several
different ways to implement it.
The system we want to describe is the particle beam
endowed with some measure of time-reversal invariance.
For such a physical situation one is faced with a many-body
dynamical system which is at the same time stochastic (due to the
random collisions and the fast, irregular configurational
dynamics) and conservative (time-reversal invariant). One then
needs a generalization of classical mechanics for conservative
deterministic systems to a classical mechanics of conservative
stochastic systems, i.e. a stochastic mechanics.

Despite the rather widespread misconception that stochastic
processes in dynamics are always associated to the description of
dissipative and irreversible behaviors, it is in fact possible for
specific stochastic dynamical systems to exhibit time-reversal
invariance. In general, these systems are such that a
deterministic dynamics in terms of known external potentials is
ascribed to a system whose kinematics, due to some intrinsic
irregularities, is instead stochastic \cite{paul}. This is the
case we conside for the collisional beam.

In previous papers~\cite{cufaropre,demartino} we
analysed some basic properties of stochastic mechanics
(SM) and we also introduced some basic criteria
of mechanical stability in order to provide
phenomenological support to the scheme of SM
for classical stochastic dynamical systems.
In particular, in Ref.~\cite{cufaropre}
we applied SM to the description of the
dynamics of charged particle beams, and
we showed how the criteria of mechanical stability
allow to connect the (transverse) emittance to the
characteristic microscopic scales of the system
and to the total number of particles in a given beam.
Moreover, this method allows
to implement techniques of active control
for the dynamics of the beam.
In Ref.~\cite{cufaropre} these techniques
have been proposed to improve the beam focusing and to
independently change the frequency of the betatron oscillations.

In the present paper, the SM of particle beams
is used to investigate
the nature, the size and the
dynamical characteristics of beam halos.
In particular, we determine the effects of the
space--charge interaction on the formation of the
halo by calculating the growth of the beam emittance,
and by estimating the probability to find
particles far away from the beam core. We then consider
the reverse problem i.e.,
by imposing reasonable forms of the stationary
halo distributions we derive the effective spatial potentials
that are associated to such distributions.

The paper is organized as follows: in Sections \ref{nelsonism}
and \ref{control} we summarize the results of previous papers
on the model of SM for particle beams
with special emphasis on the techniques of active
control needed to engineer a desired beam dynamics.
We then apply SM to study the formation of the beam halo.
In Section \ref{selfconsistent} we show how to treat with
SM the effects of space charge due to the electromagnetic
interactions among the particles.
We introduce and solve
self-consistently a set of coupled, nonlinear
 hydrodynamic equations for
the beam density and phase, the space-charge potentials, and the
external electromagnetic field. Solution of the coupled equations
leads to the first main result of our paper, i.e. that the final
transverse particle distribution is greatly broadened due to the
presence of the space-charge potential. This broadening gives rise
to a penetration of the distribution well into spatial regions
very far from the core of the beam, thus realizing a type of
continuous halo, i.e. without intermediate voids of particles
(nodes of the transverse beam distribution). This effect is
reflected in a corresponding growth of the transverse emittance
that we estimate from the solution of the self-consistent
equations. In Section \ref{halodistr} we study the reverse
problem: given different stationary transverse distributions of
the halo around the core of the beam, such as a continuous
broadened distribution or a distribution with a gap (void) between
the halo and the core, we determine the dynamics that leads to
such configurations. This is important since it enables us to
deduce the form of the effective potentials which control the
stationarity of the halo distribution. Finally, in Section
\ref{conclusions} we draw our conclusions and discuss future
directions of research.

\section{Stochastic beam dynamics}\label{nelsonism}

The usual way to introduce a stochastic dynamical system is to
modify the dynamics in phase space by adding a Wiener noise only
in the equation for the momentum, in order to preserve the usual
relations between position and velocity:
\begin{eqnarray}
m\,d{\bf Q}(t)&=&{\bf P}(t)\,dt \nonumber \\
d{\bf P}(t)&=&{\bf F}(\mathbf{Q}(t),\mathbf{P}(t),t)\,dt+\beta\,
d{\bf B}(t)
\end{eqnarray}
where $m$ is the mass of the system, $\mathbf{Q}(t)$ is the
position variable, $\mathbf{P}(t)$ is the momentum variable,
$\mathbf{F}$ is the external force, and $\mathbf{B}(t)$ is a
fluctuating force modelled by a Wiener process; in general, this
procedure yields a differentiable, but non Markovian process ${\bf
Q}(t)$ for the position (here $\beta$ denotes the autocorrelation
of the fluctuating force). The standard example of this approach
is that of a Brownian motion in a force field described by the
Ornstein--Uhlenbeck system of stochastic differential equations
(SDE) \cite{nelson}:
\begin{eqnarray}
m\,d{\bf Q}(t)\!\!&=&\!\!{\bf P}(t)\,dt \nonumber \\
d{\bf P}(t)\!\!&=&\!\!{\bf K}(\mathbf{Q}(t),t)\,dt-m\gamma
\mathbf{P}(t)\,dt+\beta\, d{\bf B}(t)
\end{eqnarray}
where $\mathbf{K}$ is the spatial gradient of the external
potential, and $\gamma$ is the damping constant. This system can
be reduced to yield a non differentiable, Markovian process ${\bf
Q}(t)$ acted on by a Wiener noise, with diffusion coefficient $D$,
to the equation for the position. In this case we are forced to
drop the equation for the momentum since ${\bf Q}(t)$ is no longer
differentiable, and we can reduce the stochastic system to the
single SDE
\begin{equation}\label{ito}
d{\bf Q}(t)\;=\;{\bf v}_{(+)}({\bf Q}(t),t)\,dt+\sqrt{D}\,d{\bf
W}(t)
\end{equation}
where ${\bf v}_{(+)}(\mathbf{r}, t)$ is the so called forward
velocity, and $d{\bf W} (t) \equiv {\bf W} (t + dt) - {\bf W} (t)$
is the increment of a standard Wiener noise $\mathbf{W}(t)$; as it
is well known, this increment is a Gaussian process.
The standard
example of this reduction is the Einstein-Smoluchowski
approximation of the Ornstein--Uhlenbeck process in
the overdamped case.

It is however possible to introduce Eq.\refeq{ito} in a completely
different context, as the defining equation for a theory of
stochastic dynamical systems in configuration space, rather than
in phase space. This amounts to consider a system whose
kinematics, rather than its dynamics, is taken {\it ab initio} to
be random. By doing this, no external sources of dissipation and
irreversibility are introduced on the forces acting on the system.
Instead, a source of randomness is assumed that perturbs only the
configurations of the system. For such systems with random
kinematics, the dynamics can then be introduced either by
generalizing the Newton equation \cite{nelson,guerraphysrep}, or
by introducing a stochastic variational principle
\cite{guerravariaz,nelson}. The crucial aspect is that such a
dynamics turns out to be both stochastic {\it and} conservative.
In this scheme, since $\mathbf{Q}(t)$ is not differentiable, there
is no velocity as a standard derivative. Instead it is possible to
define in a suitable way \cite{nelson} an average velocity ${\bf
v}_{(+)}$ in the forward time direction, and an average velocity
in the backward time direction ${\bf v}_{(-)}$: these functions of
$\mathbf{r}$ and $t$ are different, and they both coincide with
the usual velocity only if the process is differentiable, i.e. if
the kinematics is deterministic. The relations between ${\bf
v}_{(+)}$ and ${\bf v}_{(-)}$ will be introduced in the subsequent
discussion. It is important to remark that ${\bf v}_{(+)}({\bf
r},t)$ is no more an a priori given field: it plays now the role
of a dynamical variable.

The stochastic dynamical scheme sketched above is known as
Stochastic Mechanics (SM), and most of its applications have
concerned the problem of developing a classical stochastic model
for the simulation of Quantum Mechanics. Nonetheless, it is a very
general model which can be applied to very different stochastic
dynamical systems endowed with time-reversal invariance
\cite{albeverio}. We will show below how one can derive from the
stochastic variational principle two coupled, nonlinear
hydrodynamic equations for the density and the phase of a
dynamical system in configuration space. These two real, coupled
equations can be recast into a single complex equation whose form
is completely equivalent to that of a Schr\"odinger differential
equation. In this sense, some authors denote classical dynamical
systems described by SM as quantum-like systems, in analogy with
other recent studies on the collective dynamics of charged particl
beams \cite{fedele,pusterla}.

In general, SM can be used to describe every conservative,
stochastic dynamical system satisfying fairly general conditions:
for instance, it is known that for any given conservative
diffusion there is a correspondence between the associated
diffusion process and a solution of the Schr\"odinger equation
with Hamiltonians associated to suitable vector and scalar
potentials \cite{morato}. Under some regularity conditions this
correspondence is one-to-one. The usual Schr\"odinger equation,
and hence quantum mechanics, are recovered when the diffusion
coefficient $D$ is independent from the values taken by the
conservative diffusion process and coincides with $\hbar/2m$,
where $\hbar$ denotes the Planck constant. Here we are not
interested in the stochastic modelling of quantum mechanics, but
rather in the classical stochastic dynamics of particle beams.
In this instance the unit of action appearing in the effective
Schr\"odinger equation is linked to the beam emittance as will be
clarified below.

We will now describe the stochastic process performed by a
representative particle of the beam that oscillates, in a comoving
reference frame, around the closed ideal orbit in a particle
accelerator. We consider the three-dimensional (3d) diffusion
process ${\bf Q} (t)$, taking the values $\mathbf{r}$, which
describes the motion of the representative particle and whose
probability density coincides with the particle density of the
beam. The evolution of this process is ruled by the It\^o
SDE\refeq{ito} where the diffusion coefficient $D$ is supposed to
be constant. The quantity $\alpha=2mD$, which has the dimensions
of an action, will be connected later to the characteristic
transverse emittance of the beam. Eq.\refeq{ito} defines the
random kinematics performed by the collective degree of freedom,
and replaces the simple deterministic kinematics
\begin{equation}\label{classicalkin}
 d{\bf q}(t) = {\bf v}(\mathbf{q}(t),t) dt
\end{equation}
for the differentiable trajectory
$\mathbf{q}(t)$.

We are in a situation in which we have both a random, diffusive
kinematics and a time reversal invariance. Therefore, to
counteract the dissipation, one must impose a conservative
dynamics on the stochastic kinematics, at variance with the purely
dissipative Fokker-Planck or Langevin dynamics. A conservative
dynamics imposed on a random kinematics can be introduced by a
suitable stochastic generalization of the least action principle
of classical mechanics \cite{guerravariaz}. This is achieved by
replacing the classical deterministic kinematics
(\ref{classicalkin}) with the random diffusive kinematics
(\ref{ito}) as the independent configurational variables of the
Lagrangian density entering the action functional. The equations
of motion thus obtained by minimizing the action functional take
the form of two coupled hydrodynamic equations that describe the
dynamical evolution of the beam profile and of the associated
velocity field. In the following we briefly sketch the derivation
of the hydrodynamic equations, referring for details to references
\cite{guerravariaz,nelson}.

Given the SDE (\ref{ito}), we consider the probability density
function (pdf) $\rho({\bf r}, t)$ associated to the diffusion
${\bf Q}(t)$ so that, besides the forward velocity ${\bf
v}_{(+)}({\bf r}, t)$, we can also define the backward velocity
\begin{equation}\label{vback}
 {\bf v}_{(-)}({\bf r}, t)  \equiv {\bf
v}_{(+)}({\bf r}, t) - 2 D \frac{\nabla \rho ({\bf r}, t)}{\rho
({\bf r}, t)}\,.
\end{equation}
It is also useful to introduce the current and the osmotic
velocity fields, defined as:
\begin{equation}\label{osmotic}
{\bf v} \equiv \frac{{\bf v}_{(+)} + {\bf v}_{(-)}}{2} \; ;\qquad {\bf
u} \equiv \frac{{\bf v}_{(+)}-{\bf v}_{(-)}}{2}\, =\, D\frac{\nabla
\rho}{\rho} \, .
\end{equation}
The velocities in Eq.\refeq{osmotic} have a transparent physical
meaning: the current velocity ${\bf v}$ represents the global
velocity of the density profile, being the stochastic
generalization of the velocity field of a classical perfect fluid.
On the other hand the osmotic velocity $\bf u$ is clearly of
intrinsic stochastic nature, for it is a measure of the non
differentiability of the stochastic trajectories, and it is
related to the spatial variations of the density. In the limit of
a deterministic process, i.e. of a diffusion that tends to a
deterministic, differentiable trajectory ${\bf q}(t)$, the current
velocity ${\bf v}({\bf r},t)$ tends to the classical velocity
field ${\bf v}({\bf q}(t),t)$, and the osmotic velocity ${\bf u}$
tends to zero (the trajectory becomes differentiable, so that the
forward and backward velocities, i.e. the left and right
derivatives coincide).

In order to establish the stochastic generalization of the least
action principle one introduces a mean classical action
averaged over the probability density function of the
diffusion process, in strict analogy to the classical
deterministic action. In fact, the main difficulty
in the stochastic case is due to the non differentiable character
of the sample paths of a diffusion process which does not allow to
define the time derivative of the process ${\bf Q}(t)$. Hence
the definition of a Lagrangian density and of an action
functional is possible only in an average sense through a
suitable limit on expectations. The stochastic action is then
defined as \cite{guerravariaz,nelson}
\begin{equation}\label{lagrangianaction}
\mathcal{A} = \int_{t_{0}}^{t_{1}}
\lim_{\Delta t \rightarrow 0^{+}} \mathbb{E} \left[ \frac{m}{2}
\left( \frac{\Delta \mathbf{Q}}{\Delta t} \right)^{2} -
V(\mathbf{Q}) \right] dt \, ,
\end{equation}
where $\mathbb{E}(\,\cdot\,) = \int (\,\cdot\,) \rho ({\bf r},
t)\,d\mathbf{r}$ denotes the expectation of a function
of the diffusion process ${\bf Q}(t)$, $V$ is an external
potential, and $\Delta {\bf Q}(t) = {\bf Q}(t + \Delta t) - {\bf
Q}(t)$ is the finite increment of the process.
It can be shown that the mean
action (\ref{lagrangianaction}) associated to the diffusive
kinematics (\ref{ito}) can be recast in the following particularly
appealing Eulerian hydrodynamic form \cite{nelson}:
\begin{equation}\label{eulerianaction}
\mathcal{A} =
\int_{t_{0}}^{t_{1}}dt\int d{\bf r} \left[ \frac{m}{2} \left( {\bf
v}^{2} - {\bf u}^{2} \right) - V \right] \rho({\bf r}, t) \, ,
\end{equation}
where ${\bf v}$ and ${\bf  u}$ are defined in Eq.\refeq{osmotic}.
This Eulerian form of the action immediately shows that when
$\mathbf{u}=0$ (and hence the trajectories are differentiable) it
coincides with the action for the conservative dynamics of a
classical Liouville fluid. The stochastic variational principle
now follows by imposing the stationarity of the stochastic action
($\delta \mathcal{A} = 0$) under smooth and independent variations
$\delta \rho$ of the density, and $\delta {\bf v}$ of the current
velocity, with vanishing boundary conditions at the initial and
final times.

As a first consequence we get that the current velocity has the
following gradient form:
\begin{equation}\label{gradient}
m{\bf v} ({\bf r}, t) = \nabla S ({\bf r}, t) \, ,
\end{equation}
which can be also taken as the definition of the phase $S$.
The non linearly coupled Lagrange equations of motion for
the density $\rho$ and for the current velocity ${\bf v}$, of the
form (\ref{gradient}), are a continuity equation typically
associated to every diffusion process
\begin{equation}\label{continuity}
\partial_{t} \rho = -\nabla \cdot (\rho {\bf v})\,,
\end{equation}
and an evolution equation for the conservative dynamics
\begin{equation}\label{hjm}
\partial_{t} S + \frac{m}{2} {\bf v}^{2} - 2m D^2
\frac{\nabla^{2} \sqrt{\rho}}{\sqrt{\rho}} + V({\bf r}, t) = 0 \,.
\end{equation}
The above equations give a complete characterization of
time-reversal invariant diffusion processes. The last equation is
formally analogous to the Hamilton--Jacobi--Madelung (HJM)
equation originally introduced in the hydrodynamic formulation of
quantum mechanics by Madelung \cite{madelung}. However, the
physical origins of the two equations are profoundly different, as
in the quantum mechanical case the diffusion coefficient is
related to the fundamental Planck constant by the relation
$D=\hbar/2m$. Notice that when $D=0$ (namely when the kinematics
is not diffusive) the equation\refeq{hjm} coincides with the usual
Hamilton--Jacobi equation for a Liouville fluid. Since
(\ref{gradient}) holds, the two equations \refeq{continuity}
and\refeq{hjm} can be recast in the following form
\begin{eqnarray}
\partial_{t} \rho &=& -\frac{1}{m}\nabla
\cdot (\rho \nabla S)\label{continuity1}\\
 \partial_{t} S &=&- \frac{1}{2m} {\nabla S}^{2} + 2m D^2
              \frac{\nabla^{2} \sqrt{\rho}}{\sqrt{\rho}} - V({\bf
                    r},t)\label{hjm2}
\end{eqnarray}
which now constitutes a coupled, non linear system of partial
differential equations for the couple $(\rho, S)$ which completely
determines the state of our beam. On the other hand, because
of\refeq{gradient}, this state is equivalently given by the couple
$(\rho, \mathbf{v})$.

Eqns.\refeq{continuity1} and\refeq{hjm2} describe the collective
behavior of the beam at each instant of time through the evolution
of both its particle density and its velocity field. It is
important to notice that, introducing the representation
\cite{madelung}
\begin{equation}\label{schroedpsi}
\psi ({\bf r}, t) = \sqrt{\rho({\bf r}, t)}\, {\mathrm e}^{i
S({\bf r}, t)/\alpha} \, ,
\end{equation}
with
\begin{equation}
\alpha = 2mD \, ,
\label{alfa}
\end{equation}
the two coupled real equations (\ref{continuity1})
and (\ref{hjm2}) are equivalent to a single complex
Schr\"odinger equation for the function $\psi$,
with the Planck action constant $\hbar$ replaced by the
unit of action $\alpha$:
\begin{equation}\label{schroed}
  i\alpha\partial_t\psi=-\frac{\alpha^2}{2m}\nabla^2\psi+V\psi\,.
\end{equation}
In this formulation the phenomenological ``wave function'' $\psi$
carries the information on the dynamics of both the beam density
and the beam velocity field, since the velocity field is
determined via Eq.\refeq{gradient} by the phase function $S({\bf
r}, t)$. This shows, as previously claimed, that our procedure,
starting from a different point of view, leads to a description
formally analogous to that of the so-called quantum-like
approaches to beam dynamics \cite{fedele}. We remark again that
obviously Eq.\refeq{schroed} has not the same meaning as in
quantum mechanics. The parameter $\alpha$ defined in
Eq.\refeq{alfa} has the dimensions of an action, but in general
$\alpha \neq \hbar$. In fact, $\alpha$ is not a universal constant
and its value cannot coincide with $\hbar$, as its value depends
on the physical system under study. However, $\alpha$ plays in
some sense a role similar to that of $\hbar$ in quantum mechanics
since, as we will see in the following, it identifies a lower
bound for the beam emittance in phase space, and gives a measure
of the position-momentum uncertainty arising from the stochastic
dynamics of the beam. Thus the Schr\"odinger
equation\refeq{schroed} for the stochastic mechanics of particle
beams presents some features reminiscent of quantum mechanics, but
at the same time is a deeply different theory with a different
physical meaning. In particular, while in quantum mechanics a
system of $N$ particles is described by a wave function in a
$3N$--dimensional configuration space, in the scheme of SM the
normalized probability density distribution $|\psi(\mathbf{r})|^2$
is a function of only the three space coordinates in physical
space. If there are $N$ particles in the beam, then the function
$N|\psi(\mathbf{r})|^2$ is simply the particle density in physical
space.

\section{Controlled beam states}\label{control}

In the previous section we have introduced two coupled equations
that describe the dynamical behavior of the beam: the first is the
It\^o equation \refeq{ito}, or equivalently the continuity
equation (either\refeq{continuity} or\refeq{continuity1}); the
second is the Hamilton-Jacobi-Madelung (HJM) equation
(either\refeq{hjm} or\refeq{hjm2}). Here we briefly summarize a
general procedure, already exploited in
Refs.~\cite{cufarojpa,qabp1,cufaropre}, to engineer a controlled
dynamics of stochastic systems. We will then apply this method to
the control of the beam halo in Section\myref{halodistr}.

To this end it is useful to show, by simple substitution from
(\ref{osmotic}), that Eq.\refeq{continuity} is equivalent to the
Fokker--Planck (FP) equation
\begin{equation}\label{fp}
\partial_{t} \rho = -\nabla \cdot [{\bf v}_{(+)} \rho]
+  D \, \nabla^2  \rho ,
\end{equation}
formally associated to the It\^o equation\refeq{ito}. The HJM
equation\refeq{hjm} can also be cast in a form based on
$\mathbf{v}_{(+)}$ rather than on $\mathbf{v}$, namely:
\begin{eqnarray}\label{hjm1}
 \partial_t S &=&-\frac{m}{2}\mathbf{v}_{(+)}^2
     + m D\mathbf{v}_{(+)}\frac{\nabla\rho}{\rho}\nonumber\\
 &&\;\;\;\;\;\;\;\;\; +mD^2 \left[\frac{\nabla^2\rho}{\rho}
 -\left(\frac{\nabla\rho}{\rho}\right)^2\right]-V\nonumber\\
 &=& -\frac{m}{2}\mathbf{v}_{(+)}^2
        +m D\,\mathbf{v}_{(+)}\nabla\ln f\nonumber\\
 &&\;\;\;\;\;\;\;\;\;\;\;\;\;\;\;\;\;\;  +mD^2\nabla^2\ln f - V \, ,
\end{eqnarray}
where we have defined the function $f \equiv \rho/{\cal{N}}$ , and
${\cal{N}}$ is a constant such that $f$ is dimensionless. On the
other hand, we know from Eqns.\refeq{vback} and\refeq{osmotic}
that also the forward velocity $\mathbf{v}_{(+)}$ is irrotational:
\begin{equation}\label{gradient1}
 \mathbf{v}_{(+)}(\mathbf{r},t)=\nabla W(\mathbf{r},t) ,
\end{equation}
where the scalar functions $W$ and $S$
are connected by the relation
\begin{equation}\label{phases}
S(\mathbf{r},t)=mW(\mathbf{r},t)-mD\ln f(\mathbf{r},t) -
\theta(t) ,
\end{equation}
and $\theta$ is an arbitrary time-dependent function. Through
Eq.\refeq{gradient1}, Eqns.\refeq{fp} and\refeq{hjm1} can be cast
in a form analogous to the system of Eqns.\refeq{continuity1}
and\refeq{hjm2}. In this case, the couple of functions which
determine the state of the beam is $(\rho, W)$ (or equivalently
$(\rho,\mathbf{v}_{(+)})$).

It is worth noticing that the possibility of a
time reversal invariance \cite{guerraphysrep} is assured by
the fact that the forward velocity ${\bf v}_{(+)} ({\bf r},
t)$ is not an {\it a priori} given field, as is the case
for dissipative diffusion processes of the Langevin
type. Rather, it is dynamically determined at any instant
of time, for any assigned initial condition,
through the HJM evolution equation (\ref{hjm}).

Let us suppose now that the beam density $\rho(\mathbf{r},t)$ is
given at some time $t$. We may think for instance of an engineered
evolution from some initial density toward a final, required state
with suitable properties. In particular, we could imagine to have
a halo-ridden particle distribution in the beam that should be
steered toward a final halo-free distribution. We know that $\rho$
must be a solution of the FP equation \refeq{fp} for some forward
velocity field $\mathbf{v}_{(+)}(\mathbf{r},t)$, which we consider
here as not given a priori. Since also Eq.\refeq{gradient1} must
be satisfied, this means that -- for a given $\rho$ -- we should
find an irrotational $\mathbf{v}_{(+)}$ in such a way that the FP
equation \refeq{fp} is satisfied. In other words, we are required
to solve the partial differential equation
\begin{equation}
\nabla\cdot\mathbf{v}_{(+)}
+\frac{\nabla\rho}{\rho}\cdot\mathbf{v}_{(+)}=
D\frac{\nabla^2\rho}{\rho}-\frac{\partial_t\rho}{\rho} ,
\end{equation}
for the irrotational vector field $\mathbf{v}_{(+)}$, or
equivalently, by taking Eq.\refeq{gradient1} into account, the
second order partial differential equation for the scalar field
$W$:
\begin{equation}
 \nabla^2W+\frac{\nabla\rho}{\rho}\cdot\nabla W=
       D\frac{\nabla^2\rho}{\rho}-\frac{\partial_t\rho}{\rho}\,.
\end{equation}
This is not, in general, an easy task, but we will give here the
solutions for a few simple cases that will be useful in the
following.

In the one-dimensional (1d) case ($\rho(x,t)$), the FP equation
defined for $a<x<b$ (here we could possibly have $a=-\infty$, and
$b=+\infty$) is:
\begin{equation}
\partial_t\rho=-\partial_x[v_{(+)}\rho]+D\partial_x^2\rho
=\partial_xJ ,
\end{equation}
where $J=D\partial_x\rho-v_{(+)}\rho$ is the probability current.
It is easy to see now that the solution is
\begin{equation}
 v_{(+)}(x,t)=\frac{1}{\rho}
     \left[c(t)+D\partial_x\rho
-\int_a^x\partial_t\rho(x',t)\,dx'\right] ,
\end{equation}
where $c(t)$ is an arbitrary function. Moreover, since the
conservation of probability imposes, in particular, that $J(a)=0$,
it is easy to see that we must choose $c(t)=0$ so that finally we
have
\begin{equation}\label{1DIM}
 v_{(+)}(x,t)=D\frac{\partial_x\rho}{\rho}
      -\frac{1}{\rho}\int_a^x\partial_t\rho(x',t)\,dx'\,.
\end{equation}
Of course, in this case we also have
\begin{equation}
W(x,t)=\int_a^xv_{(+)}(x',t)\,dx'\,.
\end{equation}
This 1d solution will be useful later for a simplified study of
beam distributions when we take $x$ as one of the transverse
coordinates of the beam.

Another case of practical interest is that of a 3d system with
cylindrical symmetry around the $z$-axis (which will be supposed
to be the beam longitudinal axis). If we denote with $(r, \varphi,
z)$ the cylindrical coordinates, where $r=\sqrt{x^2 + y^2}$, we
will suppose that $\rho(r,t)$ depends only on $r$ and $t$, and
that $\mathbf{v}_{(+)}=v_{(+)}(r,t)\,\mathbf{\hat{r}}$ is radially
directed, with modulus depending only on $r$ and $t$. In this
case, it is straightforward to see that the equation for $v_{(+)}$
is
\begin{equation}
 \partial_rv_{(+)}+\left(\frac{1}{r}
+\frac{\partial_r\rho}{\rho}\right)v_{(+)}=
         D\left(\frac{\partial_r^2\rho}{\rho}
+\frac{\partial_r\rho}{r\rho}\right)-\frac{\partial_t\rho}{\rho} ,
\end{equation}
whose solution is
\begin{equation}\label{3DIM}
 v_{(+)}(r,t)=D\frac{\partial_r\rho}{\rho}
      -\frac{1}{r\rho}\int_0^r\partial_t\rho(r',t)r'\,dr'\,.
\end{equation}
In this case
\begin{equation}
W(r,t)=\int_0^rv_{(+)}(r',t)\,dr'\,.
\end{equation}

Finally, in the 3d stationary case the beam density $\rho$ is
independent of $t$. This greatly simplifies the treatment, as it
is easy to check by simple substitution, and we simply find
\begin{equation}\label{3DIMstaz}
 \mathbf{v}_{(+)}(\mathbf{r})=
D\frac{\nabla\rho(\mathbf{r})}{\rho(\mathbf{r})} ,
\end{equation}
with
\begin{equation}
 W(\mathbf{r})=D\ln f(\mathbf{r})\,.
\end{equation}
This result corresponds to the well known fact that for stationary
forward velocities the FP equation\refeq{fp} always admits
stationary solutions of the form
\begin{equation}\label{statsolut}
 \rho(\mathbf{r})={\cal{N}}\,e^{W(\mathbf{r})/D}\,.
\end{equation}
Remark however that also for particular choices of a non
stationary $\rho(\mathbf{r},t)$ it is sometimes possible to find a
stationary velocity field $\mathbf{v}_{(+)}(\mathbf{r})$ such that
the FP equation\refeq{fp} is satisfied (think for instance to the
Ornstein--Uhlenbeck non stationary solutions): this fact will be
of use in the following Sections.

Now that $\rho$ and $W$ (or equivalently $\mathbf{v}_{(+)}$) are
given and satisfy Eq.\refeq{fp}, we should also remember that they
will qualify as a good description of our system only if they also
are a solution of the dynamical problem, namely if they satisfy
the HJM equation\refeq{hjm1}. Since $W$, $\mathbf{v}_{(+)}$ and
$\rho$ are now fixed, this last equation must be considered as a
constraint relation defining an external controlling potential $V$
which, after straightforward calculations and in terms of the
dimensionless distribution $f$, turns out to be of the general
form:
\begin{eqnarray}\label{potential}
V(\mathbf{r},t)&=&m D^2\, \nabla^2\ln f + m D \,(\partial_t
\ln f + \mathbf{v}_{(+)}\cdot\nabla \ln f )\nonumber\\
&&\;\;\;\;\;\;-\frac{m}{2}\,\mathbf{v}_{(+)}^2 - m \partial_t W +
\dot\theta \, .
\end{eqnarray}
This procedure can in principle be applied also to more
complicated instances, for example to engineer a beam dynamics
that keeps the beam coherent even in the presence of aberrations,
or to engineer some desired evolution. However the solutions will
not always be available in closed form, and will in general
require some approximate treatment. Here we present the simple
analytic solutions for the three particular cases discussed above.

In the 1d case $f(x,t)$, with $v_{(+)}$ given by Eq.\refeq{1DIM},
the controlling potential is:
\begin{eqnarray}\label{potential1DIM}
 V(x,t)&=&\dot\theta+m D^2\, \partial_x^2\ln f + m D \,(\partial_t
 \ln f + v_{(+)}\partial_x \ln f )\nonumber\\
 &&\:\:\:-\frac{m}{2}\, v_{(+)}^2 - m \int_a^x\partial_t
 v_{(+)}(x',t)\,dx'.
\end{eqnarray}
In the 3d, cylindrically symmetric case $f(r,t)$, with $v_{(+)}$
given by\refeq{3DIM}, we have instead:
\begin{eqnarray}\label{potential3DIM}
V(r,t)\!\!&=&\!\!\frac{m D^2}{r}\partial_r(r\partial_r\ln f) +
m D (\partial_t
\ln f + v_{(+)}\partial_r \ln f )\nonumber\\
&&\;\;\;-\frac{m}{2}\,v_{(+)}^2 - m \int_0^r\partial_t
v_{(+)}(r',t)\,dr' + \dot\theta .
\end{eqnarray}
Finally, in the 3d stationary case $\rho(\mathbf{r})$ with
$\mathbf{v}_{(+)}$ given by Eq.\refeq{3DIMstaz}, the potential is
still time-dependent because of the presence of the arbitrary
function $\theta(t)$, and it reads:
\begin{equation}
V(\mathbf{r},t)=\dot\theta+2mD^2\,
\frac{\nabla^2\sqrt{\rho}}{\sqrt{\rho}}\,.
\end{equation}
In order to make it stationary it will be enough to require that
$\dot\theta(t)$ be a constant $E$, so that
\begin{equation}\label{potential3DIMstaz}
  V(\mathbf{r})=E+2mD^2\,\frac{\nabla^2\sqrt{\rho}}{\sqrt{\rho}}\,.
\end{equation}
In this stationary case the constant $E$ fixes the
zero of the potential energy, and the phenomenological wave
function \refeq{schroedpsi} takes the form
\begin{equation}
\psi(\mathbf{r},t)=\sqrt{\rho}\,e^{-i E t/\alpha}
\end{equation}
typical of stationary states.

\section{Self-consistent equations}\label{selfconsistent}

One of the possible mechanisms for the formation of the
halo in particle beams is that due to the unavoidable
presence of space charge effects. In this Section we
will investigate this possibility in the framework
of our hydrodynamic-stochastic model of beam dynamics.
To this end, we take into account
the space charge effects by coupling the hydrodynamic
equations of stochastic mechanics with the Maxwell
equations which describe the mutual electromagnetic
interactions between the particles of the beam. We
thus obtain a self-consistent,
stochastic magnetohydrodynamic system
of coupled nonlinear differential equations that can be
numerically solved to show the effect of the space charge.

In the following, the reference physical system
will be an ensemble of $N$ identical copies
of a single charged particle embedded in a
particle beam and subject to both an external
and a space-charge potential. In a reference frame
comoving with the beam, our system is then described by
the Schr\"odinger equation (\ref{schroedinger}), where
$\alpha$ is the unit of action (emittance) and
$\widehat{H}$ the Hamiltonian operator
which will be explicitly determined in the following. Since in
general $\psi$ is not normalized, we introduce the following
notation for its constant norm
\begin{equation}
\|\psi\|^2=\int_{\mathbf{R}^3}|\psi(\mathbf{r},t)|^2\,d^3\mathbf{r}\,,
\end{equation}
so that, if $N$ is the number of particles with individual charge
$q_0$, the space charge density of the beam will be
\begin{equation}
\rho_{sc}(\mathbf{r},t)=Nq_0\frac{|\psi(\mathbf{r},t)|^2}{\|\psi\|^2}
\, .
\end{equation}
On the other hand its electrical current density will be
\begin{equation}
 \mathbf{j}_{sc}(\mathbf{r},t)=Nq_0\frac{\alpha}{m}\,
   \frac{\mathrm{Im}\,
\{\psi^*(\mathbf{r},t)\mathbf{\nabla}
\psi(\mathbf{r},t)\}}{\|\psi\|^2} ,
\end{equation}
which vanishes when the wave function is stationary,
namely when
\begin{equation}\label{stationarywf}
\psi(\mathbf{r},t)=u(\mathbf{r})\,e^{-iEt/\alpha}\,.
\end{equation}
For a charged particle in the beam the electromagnetic field is
the superposition of the space-charge potential
$(\mathbf{A}_{sc},\Phi_{sc})$ due to the presence of $\rho_{sc}$
and $\mathbf{j}_{sc}$, plus the external potential
$(\mathbf{A}_{ext},\Phi_{ext})$, and hence the Schr\"odinger
equation takes the form
\begin{eqnarray}\label{schroedinger}
  i \alpha \partial_{t} \psi & = & \frac{1}{2mc^2}
  \left[ i \alpha c\nabla -q_{0}
  ( \mathbf{A}_{sc} + \mathbf{A}_{ext}) \right]^2 \psi\nonumber\\
  &&\;\;\;\;\;\;\;\;\;\;+q_0 \left( \Phi_{sc}+\Phi_{ext} \right) \psi \, .
\end{eqnarray}
Eq.\refeq{schroedinger} has then to be coupled with the Maxwell
equations for the vector and scalar potentials
$(\mathbf{A}_{sc},\Phi_{sc})$. Since we are in a reference frame
comoving with the beam, we can always assume that the wave
function is of the stationary form\refeq{stationarywf}. In this
case we have $\mathbf{j}_{sc}=0$, so that $\mathbf{A}_{sc}=0$.
Finally, by taking also $\mathbf{A}_{ext}=0$, our system is
reduced to two coupled, nonlinear equations for the pair
$(u,\Phi_{sc})$. For details see Appendix\myref{coupled}. If the
beam with space-charge interactions stays cylindrically symmetric
the function $u$ will depend only on the cylindrical radius $r$
and our system becomes (see equations\refeq{selfconstat1}
and\refeq{selfconstat2} in Appendix\myref{coupled}):
\begin{eqnarray}
 &&\!\!\!\!\!\!\!\!Eu=-\frac{\alpha^2}{2m}\left(u''
+\frac{u'}{r}\right)+
      \left(V_{ext} + V_{sc}\right)u\, ,
\label{cylindricalsc1}\\
&&\nonumber\\
 &&\!\!\!\!\!\!\!\!\frac{Nq_0^2}{2\pi\epsilon_0LA}\,u^2=
-\left(V''_{sc}+
 \frac{V'_{sc}}{r}\right)\,,\label{cylindricalsc2}
\end{eqnarray}
where
\begin{equation}
A=\int_0^{\infty}ru^2(r)\,dr \, ,
\; \; \; \|u\|^2=2\pi LA\,,
\end{equation}
with $L$ the length of the beam, and
\begin{equation}
V_{ext}(\mathbf{r},t)=q_0\Phi_{ext}(\mathbf{r},t)\, ,
\; \; \;
V_{sc}(\mathbf{r},t)=q_0\Phi_{sc}(\mathbf{r},t)\,.
\end{equation}
Eq.\refeq{cylindricalsc1} is the stationary Schr\"odinger equation
in the external scalar potential, and Eq.\refeq{cylindricalsc2} is
the Poisson equation for the space-charge scalar potential. We
proceed to solve numerically this nonlinear system and compare its
solution with that for a purely external potential $V_{ext}$ which
is a cylindrically symmetric, harmonic potential with a proper
frequency $\omega$ in absence of space charge (see
Appendix\myref{cylho}): $V_{ext} = (m\omega^{2}r^{2})/2$, with
$r=\sqrt{x^2 + y^2}$. Let us first introduce the dimensionless
quantities
\begin{eqnarray}
  s&=&\frac{r}{\sigma\sqrt{2}} \, , \nonumber \\
  &&\nonumber\\
  w(s)&=&w\left(\frac{r}{\sigma\sqrt{2}}\right)
  \equiv \sigma^{3/2}u(r) \, ,  \\
  &&\nonumber\\
  v(s)&=&v\left(\frac{r}{\sigma\sqrt{2}}\right) \equiv
  \frac{4m\sigma^2}{\alpha^2}V_{sc}(r) \, ,\nonumber
\end{eqnarray}
where $\sigma^{2} = \alpha/2m\omega$ is the variance of the ground
state of the cylindrical harmonic oscillator without space charge.
Eqns.\refeq{cylindricalsc1} and\refeq{cylindricalsc2} can then be
recast in the dimensionless form
\begin{eqnarray}
  &&\!\!\!\!\!\!\!\!s w''(s)
+w'(s)+[\beta-s^2-v(s)]\,s w(s)=0 \, , \label{adimsc1}\\
&&\nonumber\\
  &&\!\!\!\!\!\!\!\!s v''(s)+v'(s)+b\,s w^2(s)=0 \, ,
\label{adimsc2}
\end{eqnarray}
where
\begin{eqnarray}\label{parametri}
  \beta&=&\frac{4m\sigma^2}{\alpha^2}\; \;
E=\frac{2E}{\alpha\omega} \, , \nonumber \\
&&\nonumber\\
b&=&\frac{4m\sigma^2}{\alpha^2}\; \;
\frac{Nq_0^2}{2\pi\epsilon_0L}\,\frac{1}{B}=
\frac{2}{\alpha\omega}\; \;
\frac{Nq_0^2}{2\pi\epsilon_0L}\,\frac{1}{B} \, ,
 \\
&&\nonumber\\
B&=&\frac{A\sigma}{2}=\int_0^{\infty}sw^2(s) \, ds \, .\nonumber
\end{eqnarray}
The effect of the space charge will be finally accounted for by
comparing the normalized solution $w(s)$ with the unperturbed
ground state of the cylindrical harmonic oscillator
(see Appendix \myref{cylho}):
\begin{equation}\label{unpertground}
 \psi_{000}(r)=\frac{e^{-r^2/4\sigma^2}}{\sigma\sqrt{2\pi L}}\,.
\end{equation}
We have solved numerically the
system\refeq{adimsc1},\refeq{adimsc2} by tentatively fixing one of
the two free parameters $b$ and $\beta$, and then searching by an
iterative trial and error method a value of the other such that
the solution shows, in a given interval of values of $s$, the
correct infinitesimal asymptotic behavior for large values of $s$.
We have then normalized the solutions $w(s)$ by calculating
numerically the value of $B$. It is clear from the definition of
the dimensionless parameters $B$, $b$ and $\beta$
(\ref{parametri}) that the value of $\beta$ is a sort of reduced
energy eigenvalue of the system, while the product
\begin{equation}
\gamma=Bb \, ,
\label{gamma}
\end{equation}
which is by definition a non negative number, will play the role of
the interaction strength, since it depends on the space-charge
density along the linear extension of the beam.
Reverting to dimensional quantities, since
$\alpha^2/2m\sigma^2$ has the dimensions of an energy, the two
relevant parameters are
\begin{equation}\label{parameters}
E=\beta\,\frac{\alpha^2}{4m\sigma^2}\,,\qquad
\frac{Nq_0^2}{2\pi\epsilon_0L}=\gamma\,
\frac{\alpha^2}{4m\sigma^2} \, ,
\end{equation}
which are respectively the energy of the individual particle
embedded in the beam, and the strength of the space-charge
interaction.

In the following we will limit ourselves to discuss solutions of
Eqns.\refeq{adimsc1}, and\refeq{adimsc2} without nodes (a sort of
ground state for the system). It is possible to see that no
solution without nodes can be found for values of the space-charge
strength $\gamma$ beyond about $22.5$ and that for values of
$\gamma$ ranging from $0$ to $22.5$, the energy $\beta$ decreases
monotonically from $2.0000$ to $-0.0894$. If the unit of action
$\alpha$ is fixed at a given value, it is apparent that the value
of $\gamma$ is directly proportional to the charge per unit length
$Nq_0/L$ of the beam: a small value of $\gamma$ means a rarefied
beam; a large value of $\gamma$ indicates that the beam is
intense. A halo is supposed to be present in intense beams, while
in rarefied beams the behavior of every single particle tends to
be affected only by the external harmonic potential.

\begin{figure}
\begin{center}
\includegraphics*[width=8cm]{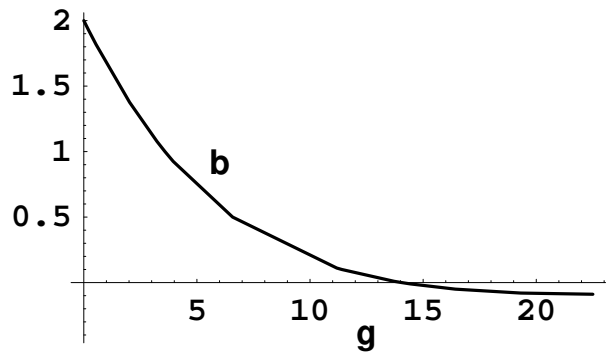}
\end{center}
\caption{The energy $\beta$ as a function of the space-charge
strength $\gamma$. Actually these quantities are dimensionless,
measured in units of the energy
$\alpha^2/4m\sigma^2\approx37.5\; \mathrm{eV}$.}\label{Bbbeta}
\end{figure}

\begin{figure}
\begin{center}
\includegraphics*[width=8cm]{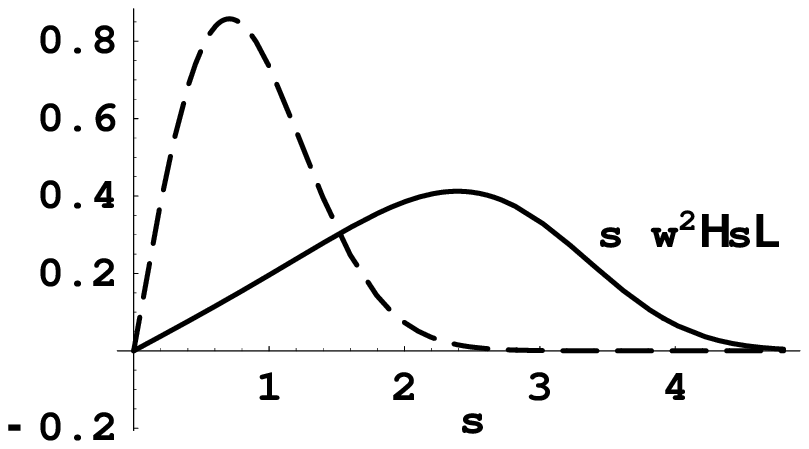}
\end{center}
\caption{The radial distribution $s\,w^2(s)$ compared with the
distribution in absence of space charge (dashed line). The
space-charge strength is $\gamma\approx22.5$. All quantities are
dimensionless.}\label{w22}
\end{figure}

\begin{figure}
\begin{center}
\includegraphics*[width=8
cm]{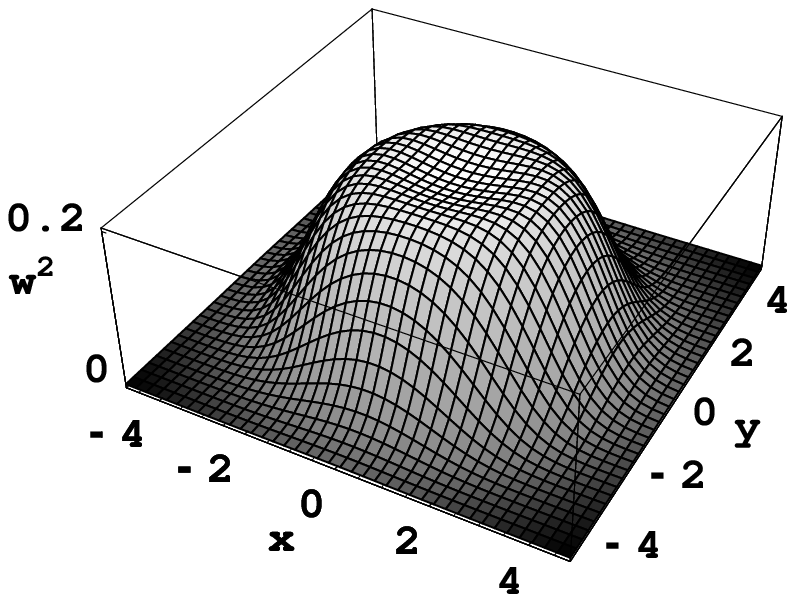}
\end{center}
\caption{3-dimensional view of the transverse distribution
$w^2(\sqrt{x^2+y^2})$ for a space-charge strength of
$\gamma\approx22.5$.}\label{w223D}
\end{figure}

\begin{figure}
\begin{center}
\includegraphics*[width=8cm]{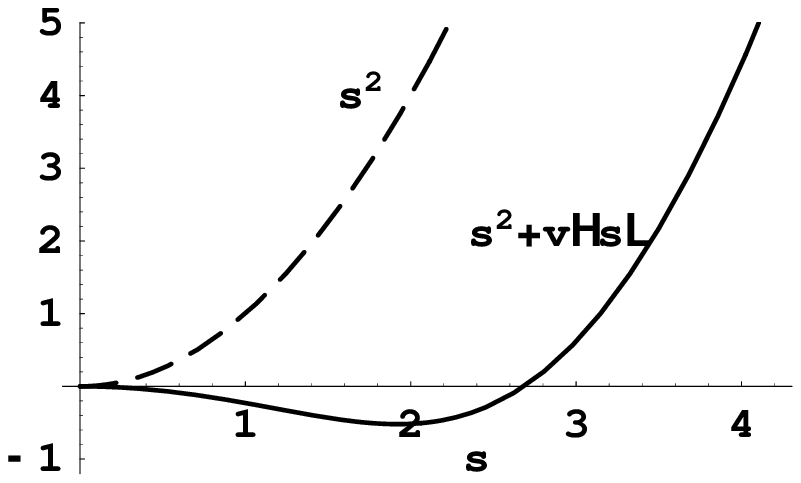}
\end{center}
\caption{The total radial potential felt by a particle in the beam
compared with the harmonic potential (dashed line) for a space
charge strength of $\gamma\approx11.2$.}\label{scpotential11}
\end{figure}

\begin{figure}
\begin{center}
\includegraphics*[width=8cm]{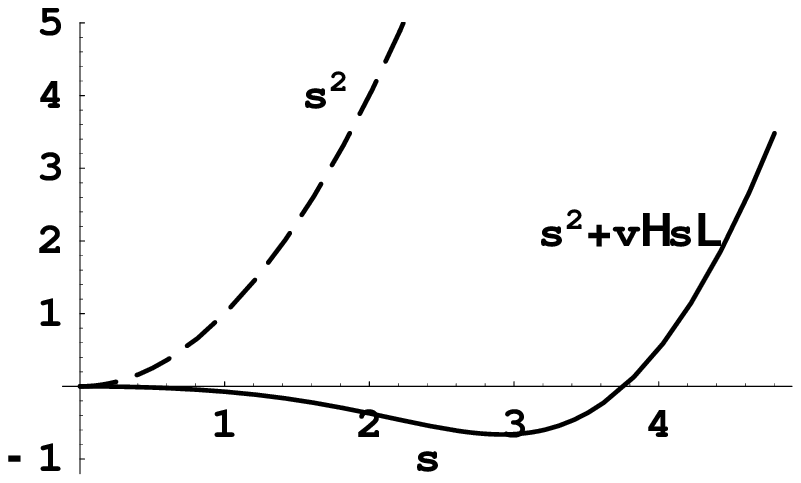}
\end{center}
\caption{The total radial potential felt by a particle in the beam
compared with the harmonic potential (dashed line) for a space
charge strength of $\gamma\approx22.5$.}\label{scpotential22}
\end{figure}

It is useful to provide a numerical estimate (in $MKS$ units) of
the relevant parameters. We are assuming a beam made of protons,
so that $m$ and $q_0$ are the proton mass and charge, while
$\epsilon_0$ is the vacuum permittivity. The parameter $\sigma$ is
determined by the strength of the external potential, and it is a
measure of the transverse size of the beam when no space charge is
taken into account, so that the ground state has the unperturbed
form\refeq{unpertground}. From empirical data, a reasonable
estimate yields $\sigma \approx 10^{-3} \;\mathrm{m}$. On the
other hand the value of $N/L$ clearly depends on the particular
beam we are considering: usually a value of $10^{11}$ particles
per meter is considered realistic. As for the parameter
$\alpha$\refeq{alfa} we have already suggested in
Section\myref{introduction} that its value can be connected to the
beam emittance. On the ground of accepted experimental values of
the beam emittance (usually measured in units of length) we can
assume $\alpha/mc$ of the order of $10^{-7} \mathrm{m}$. In fact,
taking $m$, $q_0$, $\sigma$, and $N/L$ fixed at the
above-mentioned values, and assuming a space-charge strength
$\gamma$ moderately large, i.e. $2.0 \leq \gamma \leq 22.5$, we
can determine $\alpha$ from \eqref{parameters} and we get
$7.2\times10^{-7}\leq\alpha/mc\leq1.3\times10^{-7}$, a number
which is in good agreement with the experimentally measured values
of the beam emittance. From now on we will fix $\alpha$ at its
approximate central value:
\begin{equation}
\frac{\alpha}{mc}\approx 4.0\times10^{-7}\;\mathrm{m}\, .
\end{equation}
Since $m$ and $q_0$ have the proton values, and $\sigma\approx
10^{-3}\; \mathrm{m}$, then the assigned values of the
dimensionless parameters $\gamma$ and $\beta$ will fix
respectively, the values of $N/L$ and $E$ by
Eqns.\refeq{parameters}. This implies that by changing the beam
intensity (namely $N/L$ and $\gamma$) one correspondingly changes
the energy (namely $E$ and $\beta$) of the individual particle
embedded in the beam. In Fig.\myref{Bbbeta} we show the behavior
of $\beta$ as a function of $\gamma$. The quantities $\beta$ and
$\gamma$ are dimensionless: the true physical quantities
(energies) are obtained by multiplicating them by the unit of
energy
\begin{equation}
\frac{\alpha^2}{4m\sigma^2}\approx37.5\; \mathrm{eV}\,.
\end{equation}
\begin{figure}
\begin{center}
\includegraphics*[width=8cm]{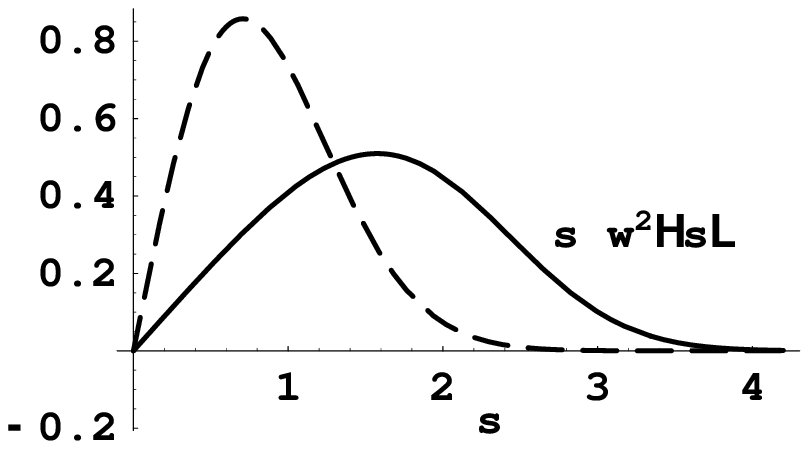}
\end{center}
\caption{The radial distribution $s\,w^2(s)$ compared with the
distribution in absence of space charge (dashed line). The
space-charge strength is $\gamma\approx11.2$. All quantities
are dimensionless.}\label{w11}
\end{figure}
\begin{figure}
\begin{center}
\includegraphics*[width=8cm]{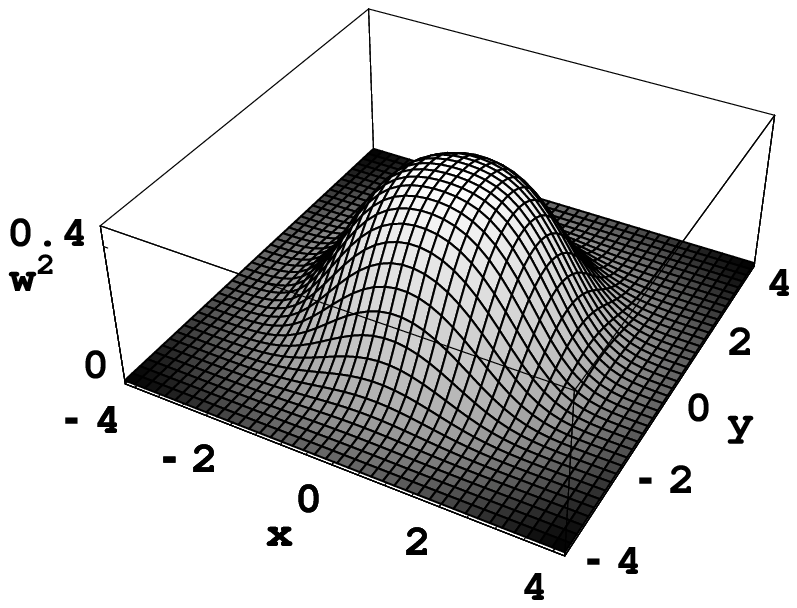}
\end{center}
\caption{3-dimensional view of the transverse distribution
$w^2(\sqrt{x^2+y^2})$ for a space-charge strength of
$\gamma\approx11.2$.}\label{w113D}
\end{figure}
The overall effect of the space charge in this model is a
conspicuous spreading of the transverse distribution of the
particles in the beam with respect to the unperturbed ground state
distribution\refeq{unpertground}. When $\gamma=0$ the potential
due to space charge vanishes, and the solution exactly coincides
with the ground state of the cylindrical harmonic oscillator with
variance $\sigma^2$. When $\gamma > 0$ the transverse distribution
begins to spread, as we show in Figrs.\myref{w11} and\myref{w113D}
for a value of the space-charge strength $\gamma\approx11.2$, and
in Figrs.\myref{w22} and\myref{w223D} for a value of the
space-charge strength $\gamma\approx22.5$. Notice that when
comparing distributions, the true radial (dimensionless) density
is $s\,w^2(s)$ and not just $w^2(s)$. Recall that $w(s)$ is a
normalized solution of Eqns.\refeq{adimsc1} and\refeq{adimsc2}. In
Figrs.\myref{scpotential11} and\myref{scpotential22} we show the
radial transverse form of the total potential, i.e. external plus
space-charge given by the solution $v(s)$ of Eqs.\refeq{adimsc1}
and\refeq{adimsc2}, for the values $\gamma=11.2$ and $\gamma=22.5$
of the space-charge strength.
To give a more quantitative measure of the flattening and
broadening of the transverse distribution we compare numerically
the probabilities of finding a particle at a relatively large
distance from the beam longitudinal axis with and without space
charge. The quantity
\begin{equation}
\mathbf{P}_{\gamma}(c)=\int_{c/\sqrt{2}}^{\infty}s\,w^2(s)\,ds
\end{equation}
is the probability of finding a particle at a
distance greater than $c\sigma$ for systems with a given
strength $\gamma$ of the space charge coupling.
For instance, considering the two different situations
$\gamma=0$ (no space charge) and $\gamma=22.5$
(strong space charge) we have:
\begin{equation}
\mathbf{P}_0(10)\approx1.9\times10^{-22} \, ,
\; \; \;
\mathbf{P}_{22.5}(10)\approx1.7\times10^{-6} \, .
\end{equation}
We see that the probability of finding particles at a distance
larger than $10\sigma$ from the core of the beam is enhanced
by space charge by many orders of magnitude. This means, for
example, that if in the beam there are $10^{11}$ particles
per meter, while practically no one is found beyond $10\sigma$ in
absence of space charge, for very strong space-charge
intensity we can find up to $10^{5}$ particles per meter at that
distance from the core.


The above analysis shows that in the hydrodynamic-stochastic
theory of charged beams, the space-charge potential induces a
strong broadening of the unperturbed transverse density
distribution of the beam, thus yielding a small, but finite
probability of having particles at a distance well away from the
core of the beam.

The hydrodynamic-stochastic model allows also an estimate on the
growth of the emittance due to the presence of the space charge.
The emittance can be calculated by exploiting a structure of
uncertainty products that is inherent to the SM. In particular,
the transverse emittance can be calculated as $\Delta x\cdot\Delta
p_x$, where we have:
\begin{equation}
\Delta x=\sqrt{\langle x^2\rangle-\langle x\rangle^2}\, ,
\; \;
\Delta p_x=\sqrt{\langle p_x^2\rangle-\langle p_x\rangle^2}
\, .
\end{equation}
More explicitly: if $w(s)$ is a solution of the coupled dynamical
equations normalized in the sense that
\begin{equation}
\int_0^{\infty}s\,w^2(s)\,ds=1\,,
\end{equation}
the corresponding 3d probability density is
\begin{equation}
\rho(r,\varphi,z)=u^2(r)H\left(\frac{L}{2}-|z|\right)\,,
\end{equation}
where $H(s)$ is the Heaviside function and
\begin{equation}
u(r)=\frac{1}{\sqrt{4\pi
L\sigma^2}}\,w\left(\frac{r}{\sigma\sqrt{2}}\right)\,.
\end{equation}
This probability density function is normalized since
\begin{equation}
\int_0^{+\infty}r \, dr
\int_0^{2\pi}d\varphi\int_{-\infty}^{\infty}dz\,
\rho(r,\varphi,z)=1\,.
\end{equation}
It is easy to show that the first two moments are:
\begin{equation}
\langle x\rangle=0\, , \; \; \langle
x^2\rangle=\sigma^2\int_0^{\infty}s^3w^2(s) \, ds
\, ,
\end{equation}
so that the root mean square deviation for the position is:
\begin{equation}
\Delta x \, = \, \sqrt{\langle
x^2\rangle} \, = \,
\sigma\sqrt{\int_0^{\infty}s^3w^2(s)\,ds} \; .
\end{equation}
As for the momentum ${\bf p}$, we recall that in SM one cannot
introduce a probability distribution directly in phase space. The
momentum can nevertheless be recovered from the velocity field
which is well defined in SM. Since we are considering a stationary
state, only the osmotic part of the velocity field is non zero.
Then the momentum field reads
\begin{equation}
\mathbf{p}(\mathbf{r})=\alpha\,
\frac{\nabla\rho(\mathbf{r})}{\rho(\mathbf{r})}
=2\alpha\,\frac{\nabla u(r)}{u(r)}\,.
\end{equation}
As a consequence, the $x$-component of the
momentum is
\begin{equation}
p_x=2\alpha\,\frac{\partial_xu}{u}=2\alpha\,
\frac{x}{r}\,\frac{u'(r)}{u(r)}=
2\alpha\cos\varphi\,\frac{u'(r)}{u(r)} \, ,
\end{equation}
and the first and second moments read
\begin{equation}
\langle p_x\rangle=0\,,\qquad\langle
p_x^2\rangle=
\frac{\alpha^2}{\sigma^2}\int_0^{\infty}{w'}^2(s)s\,ds \, .
\end{equation}
Hence the root mean square deviation for the momentum is:
\begin{equation}
\Delta p_x=\sqrt{\langle
p_x^2\rangle}=
\frac{\alpha}{\sigma}
\sqrt{\int_0^{\infty}{w'}^2(s)s\,ds} \; .
\end{equation}

\begin{figure}
\begin{center}
\includegraphics*[width=8cm]{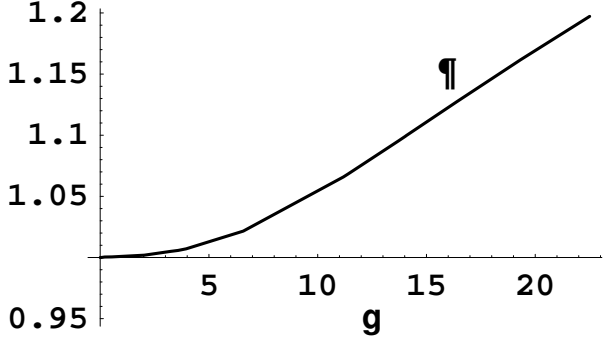}
\end{center}
\caption{The transverse dimensionless emittance $\varepsilon$ vs.
the space-charge strength $\gamma$. The true dimensional
emittance, calculated as position--momentum uncertainty product,
is just ${\cal{E}} = \varepsilon\,\alpha$.}\label{epsilon}
\end{figure}

We can then define the emittance ${\cal{E}}$
as the position-momentum uncertainty product in phase space
in the following way:
\begin{equation}
{\cal{E}} = \Delta x\cdot\Delta
p_x=\alpha \sqrt{\int_0^{\infty}s^3w^2(s) ds
\cdot\int_0^{\infty}{w'}^2(s)s ds} .
\label{emittance}
\end{equation}
The emittance (\ref{emittance}) can be thus estimated from
knowledge of the numerical solutions $w(s)$ and their first
derivatives $w'(s)$. It can be seen that the emittance is exactly
$\alpha$ for $\gamma=0$ (namely in absence of space charge), and
grows with $\gamma$ to a value $\,\approx 1.2\times\alpha$ for
$\gamma\approx22.5$. This result is consistent with the expected
growth of emittance produced by space-charge effects. The
dimensionless ratio $\varepsilon={\cal{E}}/\alpha = \Delta
x\cdot\Delta p_x/\alpha$ is plotted as a function of $\gamma$ in
Fig.\myref{epsilon}. This figure and the above discussion provide
evidence that in the model of SM of charged beams the constant
$\alpha$ plays the role of a lower bound for the phase-space
emittance.

\section{Stationary halo distributions}\label{halodistr}

The coupled nonlinear equations introduced in the previous Section
allow to introduce and explain a possible mechanism of halo
formation due to space-charge effects. This mechanism, when
considering solutions without nodes, amounts to the broadening of
the transverse density distribution of the beam away from the beam
core. This is a type of ``continuous'' halo. If we want to
investigate the formation and the properties of other possible
forms of halos -- for example ring-like halos separated from the
core by an almost void space region -- we should try to analyze
the solutions with nodes of our radial self-consistent equations.
Being this a time consuming task, we resort for the time being to
a different, simplified approach to study the structure and
properties of beams of the ringed type.

Moreover, the scheme introduced in the previous Section, although
in principle very general, has the drawback that it does not allow
for analytical expressions of velocities and potentials, and
therefore cannot suggest the engineering of controlling external
potentials to remove the halo. To this end, in this Section we
will develop a different approach, still based on SM, by
considering some preassigned density distributions for a beam with
halo, and then looking for the kind of dynamics that produces
them.

\begin{figure}
\begin{center}
\includegraphics*[width=8cm]{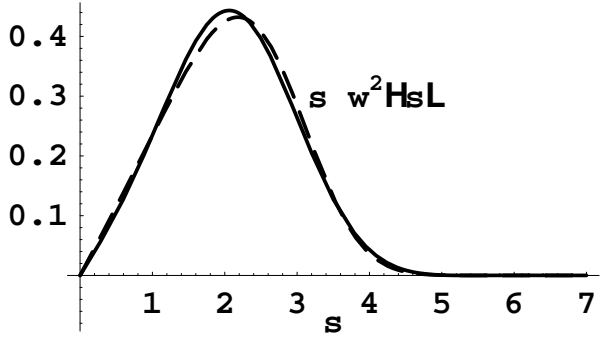}
\end{center}
\caption{The best approximation in mean square (solid line) of a
numerical solution for the self-consistent equations (dashed line)
for trial functions of the form\refeq{trialw} with
$\gamma\approx19.3$.}\label{fit01}
\end{figure}

\begin{figure}
\begin{center}
\includegraphics*[width=8cm]{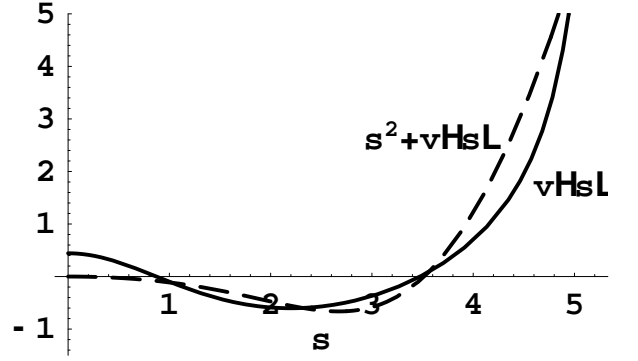}
\end{center}
\caption{The control potential (solid line) obtained from the best
mean square approximation $\widetilde{w}(s)$ to a numerical
solution for the self--consistent equations\refeq{adimsc1}
and\refeq{adimsc2}, for trial functions of the form\refeq{trialw}
with $\gamma = 19.3$. It is compared with the potential $s^2+v(s)$
(dashed line) obtained as numerical solution of\refeq{adimsc1}
and\refeq{adimsc2}.}\label{fit03}
\end{figure}

Firs of all we will determine the analytical forms
$\widetilde{w}(s)$ of the dimensionless radial distribution that
can best approximate the numerical, self--consistent solutions
$w(s)$ found in the previous Section. We will do that by choosing
a family of trial functions $\widetilde{w}(s)$ and by subsequently
minimizing the mean square error (m.s.e.) with respect to a given
solution $w(s)$. Since the radial component of the cylindrically
symmetric ($l=0$) solutions of the harmonic
oscillator\refeq{cylindrHO} contain only even powers of the radial
coordinate (see Appendix B), we take as normalized trial functions
\begin{equation}\label{trialw}
  \widetilde{w}(s)=\frac{e^{-s^2/2\sigma^2}}{\sqrt{c}}\,(1+as^2+bs^4)
\end{equation}
with $\sigma$, $a$ and $b$ as free parameters, and
\begin{eqnarray}
 c\!\!&=&\!\!\int_0^{\infty}e^{-s^2/\sigma^2}(1+as^2+bs^4)^2s\,ds
\nonumber \\
 &=&\!\!\frac{\sigma^2}{2}+a\sigma^4+(a^2+2b)\sigma^6
   +6ab\sigma^8+12b^2\sigma^{10}
\end{eqnarray}
as normalization constant. We then take one numerical solution
$w(s)$ of\refeq{adimsc1} and\refeq{adimsc2} for a given value of
$\gamma$ and consider the mean square error
\begin{equation}
J=\int_0^{\infty}|w(s)-\widetilde{w}(s)|^2s\,ds \, ,
\end{equation}
and numerically minimize its value by varying the parameters
$\sigma$, $a$ and $b$.
As an example, for $\gamma\approx19.3$, we get
\begin{equation}
\sigma\approx1.81\,;\quad a\approx0.28\,;\quad
b\approx0.01\,;\quad c\approx5.09 \, .
\label{values}
\end{equation}
The corresponding density distribution (solid line) is compared in
Fig.\myref{fit01} with the density distribution obtained by
numerically solving the system\refeq{adimsc1} and\refeq{adimsc2}
(dashed line).
From this explicit form of the approximating $\widetilde{w}(s)$ we
can now also calculate the expression of the stationary potential
which controls the state. From Eq.\refeq{potential3DIMstaz} in the
case of a cylindrically symmetric approximating state with
\begin{equation}
\widetilde{u}(r)=\sqrt{\widetilde{\rho}(r)} \, ,
\end{equation}
we have the controlling potential
\begin{equation}
\widetilde{V}(r)=\widetilde{E}+\frac{\alpha^2}{2m}
\left[\frac{\widetilde{u}\,''(r)}{\widetilde{u}(r)}+
\frac{1}{r}\frac{\widetilde{u}\,'(r)}{\widetilde{u}(r)}\right]\,.
\end{equation}
Introducing the usual dimensionless quantities
\begin{equation}
 s=\frac{r}{\sigma\sqrt{2}}\,,\quad
 \widetilde{\beta}=\frac{4m\sigma^2}{\alpha^2}\,\widetilde{E}\,;\quad
 \widetilde{w}(s)=\sigma^{3/2}\widetilde{u}(r) \, ,
\end{equation}
we may define the dimensionless controlling potential
\begin{eqnarray}
 \widetilde{v}(s)&=&
\widetilde{v}\left(\frac{r}{\sigma\sqrt{2}}\right)
\equiv \frac{4m\sigma^2}{\alpha^2}\,\widetilde{V}(r)
\nonumber \\
&& \nonumber \\
 &=&\widetilde{\beta} +
\frac{\widetilde{w}\,''(s)}{\widetilde{w}(s)}
+\frac{1}{s}\frac{\widetilde{w}\,'(s)}{\widetilde{w}(s)}
\, .
\end{eqnarray}
For an approximating amplitude of the form\refeq{trialw}, and with
the same values\refeq{values} of the parameters, the controlling
potential reads
\begin{equation}
\widetilde{v}(s)=\frac{0.442-0.629 s^2+0.062 s^4-0.001
s^6}{1+0.284 s^2-0.011 s^4} \, .
\end{equation}
This potential is shown (solid line) in Fig.\myref{fit03}, where
it is compared with the potential $s^2+v(s)$ (dashed line)
solution of Eqns.\refeq{adimsc1} and\refeq{adimsc2}.

\begin{figure}
\begin{center}
\includegraphics*[width=8cm]{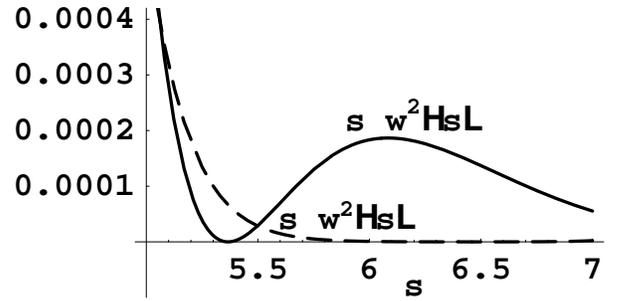}
\end{center}
\caption{Blown up picture of Fig. \ref{fit01} that shows the
existence of a node in the position density distribution at
$s\approx 5.4$.} \label{fit02}
\end{figure}

The analytical approximations show a small, but relevant,
difference with respect to the numerical solutions of
Eqns.\refeq{adimsc1} and\refeq{adimsc2}: $\widetilde{w}(s)$ has a
node at about $s\approx5.4$, and correspondingly the potential
$\widetilde{v}(s)$ has a singularity at that point (out of the $s$
range of Fig.\myref{fit03}). Instead, the numerical solutions
$w(s)$ and $v(s)$ of Eqns.\refeq{adimsc1} and\refeq{adimsc2} show
no such behavior. For values of $s$ beyond $s\approx5.4$,
$\widetilde{w}(s)$ shows a small bump, not visible in
Fig.\myref{fit01} because of the large scale. We show a zoom-up of
it in Fig.\myref{fit02}, where the node is now clearly visible.
This is a feature that we will find as well in other different
approaches to the description of the beam distribution, and that
could be tentatively connected with the presence of a halo
spatially separated from the core of the beam.

\begin{figure}
\begin{center}
\includegraphics*[width=8cm]{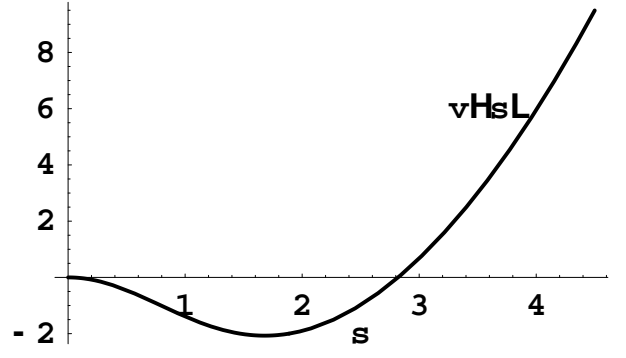}
\end{center}
\caption{The dimensionless space-charge potential $v(s)$
produced by a frozen, cylindrically symmetric beam
distribution with $\gamma=3$.} \label{frozenpotential}
\end{figure}

In our quest for an analytic description of the halo distribution
we now depart from the scheme of the self-consistent nonlinear
equations\refeq{adimsc1} and\refeq{adimsc2}. Instead, while still
assuming the space charge as the dominant factor, we introduce a
simplified analytical model for the halo production. As a first
simplification, valid for small values of the space--charge
strength $\gamma$, we assume a frozen core for the beam. A frozen
core is such that it produces a space-charge effect, induces a
spreading of the density distribution, and is left unchanged by
this spreading. In this case the potential produced by the charge
distribution is a priori given by the distribution of the frozen
core and it simply enters linearly in the phenomenological
Schr\"odinger equation\refeq{schroed}. As a consequence, the
latter is now decoupled from the Maxwell equations for the
electromagnetic field. The frozen charge distribution will be
assumed cylindrically symmetric and transversally Gaussian:
\begin{equation}
 Q(\mathbf{r})=Nq_0\frac{e^{-(x^2+y^2)/2\sigma^2}}{2\pi\sigma^2}\,
 \frac{1}{L}\,H\left(\frac{L}{2}-|z|\right) \, ,
\end{equation}
where $N$ is the number of particles with elementary charge $q_0$,
$L$ is the longitudinal extension of the beam, and $H(\cdot)$
denotes the Heaviside function. The cylindrical symmetry allows to
determine via Gauss theorem the potential energy $V_{sc}(r)$ of a
test particle of charge $q_0$ embedded in this charge
distribution. It is a simple exercise to show that
\begin{equation}
 V_{sc}(r)=\frac{Nq_0^2}{2\pi\epsilon_0L}
     \left[\mathrm{Ei}\left(-\frac{r^2}{2\sigma^2}\right)
           -\ln\frac{r^2}{2\sigma^2}-\mathbb{C}\right] \, ,
\end{equation}
where $\mathbb{C}\approx0.577$ is the Euler constant, and
\begin{equation}
\mathrm{Ei}(x)=\int_{-\infty}^x\frac{e^t}{t}\,dt
\end{equation}
is the exponential--integral function. If we suppose that the test
particle is acted upon by this space-charge potential, and by the
external, cylindrical harmonic oscillator
potential\refeq{cylindrHO} -- which, without space charge, would
keep the particle in the Gaussian state with variance $\sigma^2$
-- the total potential to be inserted in the Schr\"odinger
equation \refeq{schroed} reads
\begin{eqnarray}
 V(r)&=&\frac{\alpha^2}{8m\sigma^4}r^2\nonumber\\
 &&\nonumber\\
 &&\;\;\;\;+\frac{Nq_0^2}{2\pi\epsilon_0L}
     \left[\mathrm{Ei}\left(-\frac{r^2}{2\sigma^2}\right)
           -\ln\frac{r^2}{2\sigma^2}-\mathbb{C}\right] .
\end{eqnarray}
In terms of the usual dimensionless quantities we can also write
\begin{eqnarray}
 s&=&\frac{r}{\sigma\sqrt{2}}\,,
     \qquad\gamma\;=\; \frac{4m\sigma^2}{\alpha^2}
\frac{Nq_0^2}{2\pi\epsilon_0L} \nonumber \\
&& \nonumber \\
v(s)&=&v\left(\frac{r}{\sigma\sqrt{2}}\right)
\equiv\frac{4m\sigma^2}{\alpha^2}\,V(r) \nonumber \\
&& \nonumber \\
 &=&s^2+\gamma[\,\mathrm{Ei}(-s^2)-\ln s^2-\mathbb{C}\,]
\end{eqnarray}
A plot of the dimensionless potential $v(s)$ for $\gamma=3$ is
shown in Fig.\myref{frozenpotential}.

\begin{figure}
\begin{center}
\includegraphics*[width=8cm]{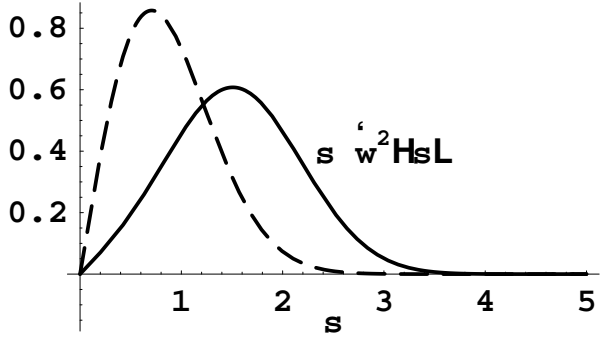}
\end{center}
\caption{The variational approximation to the density distribution
produced by the space charge of a frozen core with $\gamma=3$
(solid line). It is compared with the Gaussian solution (dashed
line) corresponding to the case of a purely harmonic potential
($\gamma=0$).} \label{frozen01}
\end{figure}

\begin{figure}
\begin{center}
\includegraphics*[width=8cm]{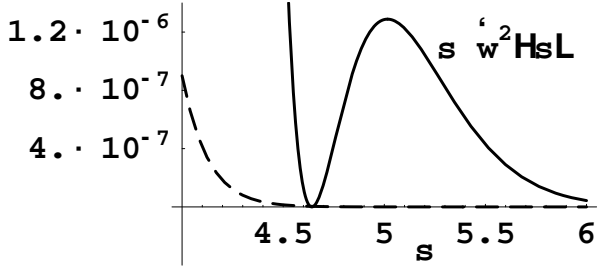}
\end{center}
\caption{Zoom up of Fig.\myref{frozen01} that shows a node in the
position density distribution at $s\approx4.7$.} \label{frozen02}
\end{figure}

\begin{figure}
\begin{center}
\includegraphics*[width=8cm]{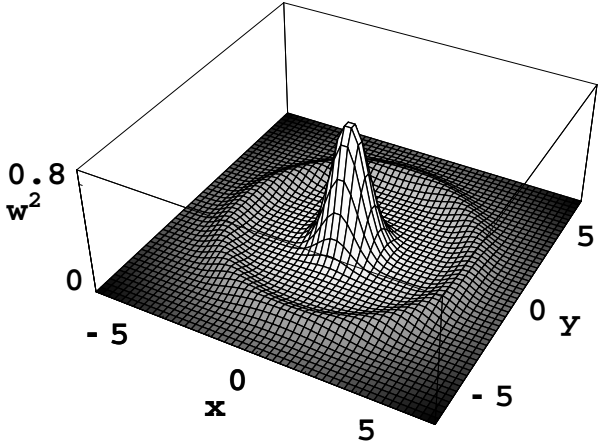}
\end{center}
\caption{Plot of the cylindrically symmetric density distribution
of the beam $w^{2}(\sqrt{x^{2}+y^{2}})$ given by
Eq.\refeq{adimensionale}, which shows a halo ring surrounding the
beam core.} \label{ring01}
\end{figure}

\begin{figure}
\begin{center}
\includegraphics*[width=8cm]{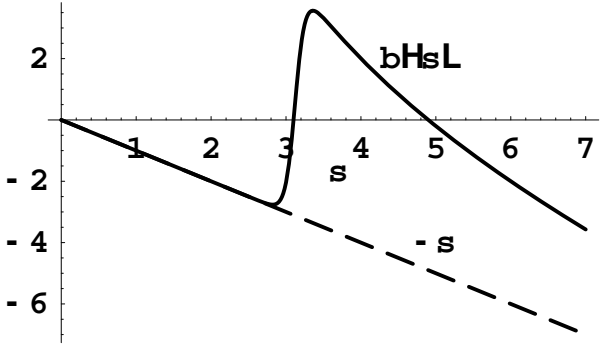}
\end{center}
\caption{The radial, dimensionless, forward velocity field for the
distribution of Fig.\myref{ring01} (solid line), and for the
ground state of a harmonic oscillator (dashed line).}
\label{ring03}
\end{figure}

\begin{figure}
\begin{center}
\includegraphics*[width=8cm]{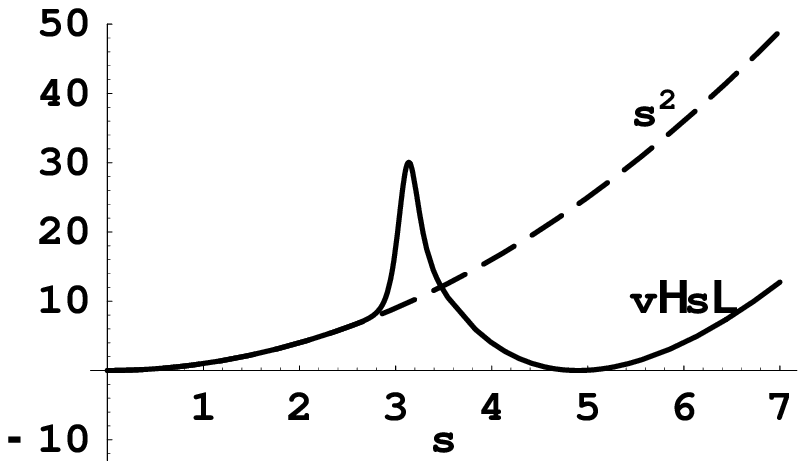}
\end{center}
\caption{The radial, dimensionless, potential for the distribution
of Figure\myref{ring01} (solid line), and for a harmonic
oscillator (dashed line).} \label{ring02}
\end{figure}

For stationary solutions of the form
\begin{equation}
\psi(\mathbf{r},t)=u(\mathbf{r})\,e^{-iEt/\alpha} \, ,
\end{equation}
Eq.\refeq{schroed} becomes
\begin{equation}
E\,u(\mathbf{r})=\widehat{H}\,u(\mathbf{r})
=-\frac{\alpha^2}{2m}\nabla^2u(\mathbf{r})+V(\mathbf{r})
\,u(\mathbf{r}) \, ,
\end{equation}
and due to the cylindrical symmetry of $V$ it reduces to
\begin{equation}
E\,u(r)=-\frac{\alpha^2}{2m}\left[u''(r)+
\frac{u'(r)}{r}\right]+V(r)\,u(r) \, .
\end{equation}
Considering the dimensionless quantities
\begin{eqnarray}
\beta&=&\frac{4m\sigma^2}{\alpha^2}\,E\,, \; \;
w(s)=w\left(\frac{r}{\sigma\sqrt{2}}\right) \equiv\sigma^{3/2}u(r)
\end{eqnarray}
we finally obtain the reduced equation
\begin{equation}
\beta w(s) = -w''(s)-\frac{w'(s)}{s}+v(s)w(s)\,.
\label{reduced}
\end{equation}
Despite its simple graphical behavior displayed in
Fig.\myref{frozenpotential} the potential $v(s)$ is complicated
enough to bar the hope of exactly solving Eq.\refeq{reduced} in a
simple way. Reasonable approximate solutions can be obtained by
means of the variational method of Rayleigh and Ritz. We thus look
for the minimum of the average energy
\begin{equation}
\frac{(u,\widehat{H}\,u)}{(u,u)} \, ,
\end{equation}
whose reduced and dimensionless form is
\begin{equation}
\frac{\int_0^{\infty}[-w(s)w''(s)
s-w(s)w'(s)+v(s)w^2(s)s]ds}{\int_0^{\infty}w^2(s)sds} \, .
\end{equation}
The minimization can be carried out for different values of the
coupling parameter $\gamma$, but the frozen core model is
realistic only for small values of $\gamma$: a strong coupling
with a frozen core would have the paradoxical effect of totally
expelling the distribution function of the test particle from the
center of the beam, while keeping the Gaussian core always
concentrated in the center. To elaborate an example we have chosen
$\gamma=3$ and a test function of the form
\begin{equation}
w(s)=\frac{e^{-s^2/2\sigma^2}}{\sqrt{c}}\,(1+as^2+bs^4) \, .
\end{equation}
The values of the parameters that minimize the energy
functional turn out to be
\begin{equation}
\sigma = 1.24\,;\quad a = 0.63\,;\quad
b = 0.03\,;\quad c = 2.92\,.
\end{equation}
The corresponding optimal radial density distribution
$s\widehat{w}\,^2(s)$ is shown in Fig.\myref{frozen01} where it is
also compared with the Gaussian solution (dashed line)
corresponding to the case of a purely harmonic potential
($\gamma=0$).
Notice that this density distribution has a form similar to that
of the self--consistent solutions elaborated in
Section\myref{selfconsistent} for values of $\gamma$ of about $10$
(see Fig.\myref{w11}). It has a node which is not visible in
Fig.\myref{frozen01}. We zoom up its plot in Fig.\myref{frozen02}.
This behavior suggests that the halo distribution could be
described, in first approximation, as a Gaussian core distribution
plus a small ring of particles surrounding it and constituting the
halo. Thus, in the following we will simply assume a specific form
of a beam with a halo without deriving it as an effect of space
charge interactions. Since it is not clearly established that a
halo can be due only to space-charge effects, starting with a
realistic ring distribution and trying to understand the dynamics
that can produce it could be very useful in this respect. The
techniques introduced in Section\myref{control} will be
instrumental to this end. For a three-dimensional, cylindrically
symmetric beam we introduce the normalized radial density
distribution
\begin{eqnarray}\label{3dimcylpdf}
\rho(r) & = & A\,\frac{e^{-r^2/2\sigma^2}}{\sigma^2}\nonumber\\
        &&\;\;\;\;+(1-A)
     \frac{e^{-r^2/2p^2\sigma^2}}{p^2\sigma^2\Gamma(q+1)}
 \left(\frac{r^2}{2p^2\sigma^2}\right)^q
\end{eqnarray}
which is composed of a Gaussian core with variance $\sigma^2$ (the
simple harmonic oscillator ground state), plus a ring-like
distribution whose size is fixed through the two parameters $p>0$
and $q\geq0$. The parameter $0\leq A\leq1$ is the relative weight
of the two parts. This density distribution has the required form
for suitable values of the parameters, but, at variance with the
two previous examples of approximate distributions (see
Figs.\myref{fit02} and\myref{frozen02}), it has no nodes. This is
convenient for two main reasons: first because it is a rather
general requirement for a ground state to have no nodes (this fact
is a rigorous theorem for one-dimensional systems). Moreover, it
has been shown in Refs.~\cite{cufarojpa,bvth,cufaropla} that
stationary distributions without nodes are also attractors for
every other possible (non extremal) initial distribution: a
property that will be useful in a future discussion of the
possible relaxation of the system toward a stable beam halo. For
this cylindrically symmetric case the expressions\refeq{3DIMstaz}
and\refeq{potential3DIMstaz} for the velocity field and the
potential become
\begin{eqnarray}
 u(r) & = & \sqrt{\rho(r)} \, , \nonumber \\
&& \nonumber \\
 v_{(+)}(r)&=&\frac{\alpha}{m}\,\frac{u'(r)}{u(r)}
\, , \nonumber \\
&& \nonumber \\
 V(r) & = & \frac{\alpha^2}{2m\sigma^2}+\frac{\alpha^2}{2m}
        \left(\frac{u''(r)}{u(r)}+\frac{1}{r}
\frac{u'(r)}{u(r)}\right) \, .
\end{eqnarray}
Moving to dimensionless quantities we have
\begin{eqnarray}
s & = & \frac{r}{\sigma\sqrt{2}} \, , \nonumber \\
w^2(s) & = & w^2\left(\frac{r}{\sigma\sqrt{2}}\right)
=2\sigma^2\rho(r) \nonumber \\
& = & 2Ae^{-s^2}+2(1-A)\frac{e^{-s^2/p^2}}{p^2\Gamma(q+1)}
\left(\frac{s^2}{p^2}\right)^q \, , \nonumber \\
&& \nonumber \\
b(s) & = & b \left(\frac{r}{\sigma\sqrt{2}}\right)
=\frac{m\sigma\sqrt{2}}{\alpha}v_{(+)}(r)
=\frac{w'(s)}{w(s)} \, ,\nonumber \\
&& \nonumber \\
v(s) & = & v\left(\frac{r}{\sigma\sqrt{2}}\right)
=\frac{4m\sigma^2}{\alpha^2}V(r) \nonumber \\
&& \nonumber \\
& = & 2+\frac{w''(s)}{w(s)}+\frac{1}{s}\,\frac{w'(s)}{w(s)}
\label{adimensionale}
\end{eqnarray}
A three-dimensional plot of the dimensionless density distribution
is given in Fig.\myref{ring01} where we have chosen $p=1$, $q=24$
and $A\approx0.49$. In this example the importance of the halo
ring has been exaggerated with respect to the real case in order
to make the effect clearly visible. The explicit analytic
expression of $b(s)$ and $v(s)$ is rather lengthy and not
particularly illuminating: we simply show the behavior of these
functions, for the same values of the parameter for the previous
example, respectively in Fig.\myref{ring03} and
Fig.\myref{ring02}, where they are also compared with the
corresponding dimensionless quantities ($-s$ and $s^2$) for the
ground state of the harmonic oscillator. The function $v(s)$ is
the potential that in the stochastic model keeps the beam in the
stationary state with a halo ring.

\begin{figure}
\vspace{0.4cm}
\begin{center}
\includegraphics*[width=8cm]{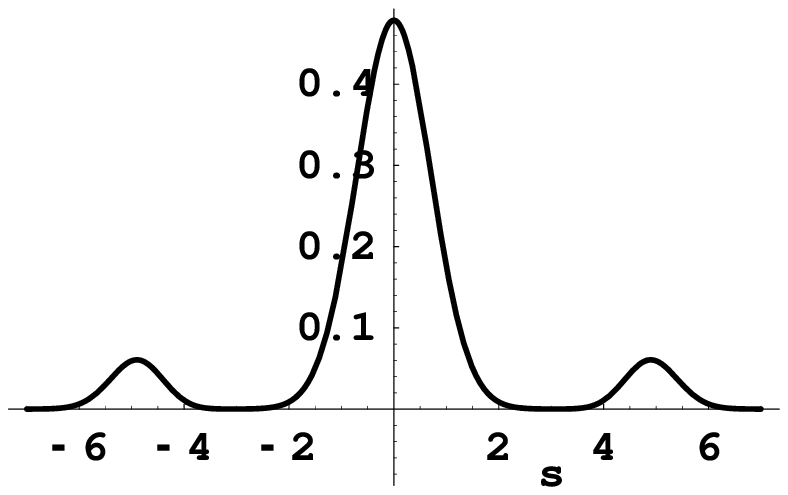}
\end{center}
\caption{Plot of the 1d density distribution\refeq{ringpdf1dim}
with a halo ring surrounding the beam core.}\label{ring1dim01}
\end{figure}

\begin{figure}
\begin{center}
\includegraphics*[width=8cm]{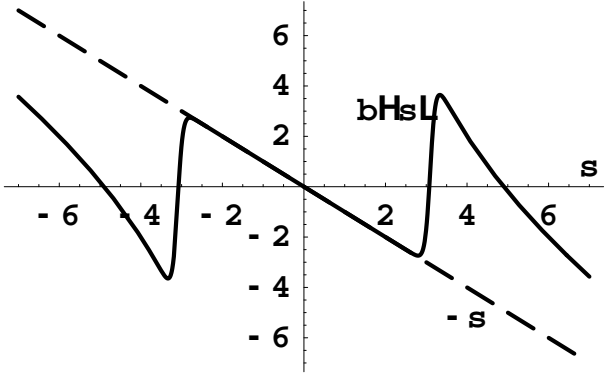}
\end{center}
\caption{The dimensionless velocity for the 1d distribution of
Figure\myref{ring1dim01} (solid line), and for a harmonic
oscillator (dashed line).}\label{ring1dim03}
\end{figure}

\begin{figure}
\begin{center}
\includegraphics*[width=8cm]{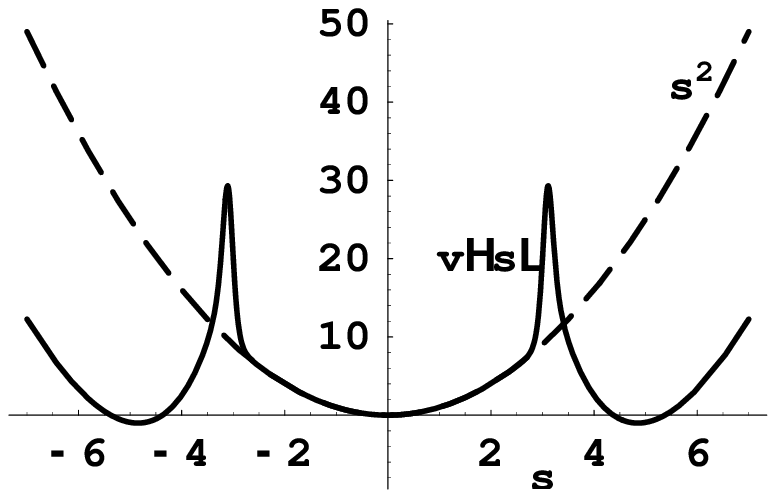}
\end{center}
\caption{The dimensionless potential for the 1d distribution of
Figure\myref{ring1dim01} (solid line), and for a harmonic
oscillator (dashed line).}\label{ring1dim02}
\end{figure}

In order to illustrate the main effects involved, let us give here
the simpler formulae relative to the 1d case. From now on we will
consider only 1d processes denoting by $x$ one of the transverse
space coordinates. We assume that the longitudinal and the
transverse beam dynamics can be deemed independent, with the
further simplification of considering decoupled evolutions along
the transverse directions $x$ and $y$. Under these conditions the
density distribution with a halo ring now reads
\begin{eqnarray}\label{ringpdf1dim}
 \rho(x)&=&A\,\frac{e^{-x^2/2\sigma^2}}{\sigma\sqrt{2\pi}}\nonumber\\
 && \;\;\;+(1-A)\frac{e^{-x^2/2p^2\sigma^2}}{p\,\sigma\sqrt{2}
 \Gamma\left(q+\frac{1}{2}\right)}
 \left(\frac{x^2}{2p^2\sigma^2}\right)^q
\end{eqnarray}
so that the corresponding velocities and potential are
\begin{eqnarray}
 u(x)&=&\sqrt{\rho(x)}\,,\nonumber\\
 v_{(+)}(x)&=&\frac{\alpha}{m}\,\frac{u'(x)}{u(x)}\,,\nonumber\\
 V(x)&=&\frac{\alpha^2}{4m\sigma^2}+\frac{\alpha^2}{2m}
        \frac{u''(x)}{u(x)}\,.\label{1dimpot}
\end{eqnarray}
In terms of dimensionless quantities we then have for the
distribution
\begin{eqnarray}
 s&=&\frac{x}{\sigma\sqrt{2}}\nonumber\\
 w^2(s) & = & w^2\left(\frac{x}{\sigma\sqrt{2}}\right)
=2\sigma^2\rho(x)\nonumber\\
& = & A\,\frac{e^{-s^2}}{\sqrt{\pi}}
+(1-A)\frac{e^{-s^2/p^2}}{p\,\Gamma\left(q+\frac{1}{2}\right)}\,
\left(\frac{s^2}{p^2}\right)^q \label{ringdlpdf1dim}
\end{eqnarray}
while the forward velocity and the potential are
\begin{eqnarray}
b(s)\!\!&=&\!\!b\left(\frac{x}{\sigma\sqrt{2}}\right)
=\frac{m\sigma\sqrt{2}}{\alpha}v_{(+)}(x)
=\frac{w'(s)}{w(s)}\label{ringvel1dim} \, , \\
&& \nonumber \\
v(s)\!\! & = &\!\! v \left(\frac{x}{\sigma\sqrt{2}}\right)
=\frac{4m\sigma^2}{\alpha^2}V(x) =1+\frac{w''(s)}{w(s)} \, .
\label{ringpot1dim}
\end{eqnarray}
In Figrs.\myref{ring1dim01},\myref{ring1dim03},
and\myref{ring1dim02} we respectively show the density
distribution, the velocity field, and the potential for $p=1$,
$q=24$ and $A=0.85$. As in the three-dimensional case, the
importance of the halo ring has been exaggerated with respect to
the real case in order to make the effect clearly visible.

\section{Conclusions}\label{conclusions}

In this paper we have presented a dynamical, stochastic approach
to the description of the beam transverse distribution in the
particle accelerators. In the first part we have described the
collective beam dynamics in terms of time--reversal invariant
diffusion processes (Nelson processes) which are obtained by a
stochastic extension of the least action principle of classical
mechanics. The diffusion coefficient of the process is identified
with a lower bound for the emittance of the beam as discussed in
the Section\myref{selfconsistent}. The collective dynamics of
beams is then described by two nonlinearly coupled hydrodynamic
equations. This set of equations shows some formal analogies with
the effective Ginzburg-Landau and Gross-Pitaevskii descriptions of
the collective dynamics of self-interacting quantum many body
systems in terms of classical, nonlinear Schr\"odinger equation.
We have shown that, in the framework of SM, it is possible to have
transverse distribution which show a broadening and an emittance
growth, typical of the halo formation. The interest of this
approach lies in the fact that its dynamical equations allow us,
at least in principle, to determine the possible control
potentials that are responsible for the stationarity of the halo.
A few introductory examples of these potentials and distributions
have been discussed in the section~\myref{halodistr}. The
controlling potentials can be engineered by suitable tuning of the
external electromagnetic fields. In a previous
paper~\cite{cufaropre} we have considered evolutions that drive
the beam from a less collimated to a better collimated state, and
we have furthermore shown that this goal can also be achieved
without increasing the frequency of the betatron oscillations
which can in fact be independently controlled during the
evolution. In forthcoming papers we plan to elaborate on the
extension of these techniques to the problem of engineering
suitable time-dependent potentials for the control and the
elimination of the beam halo. In particular, it will be shown that
the transition functions of the Nelson processes can be exploited
to control these evolutions. We also plan to extend the analysis
of the present paper to different, non Gaussian beam transverse
distributions which could better describe the existence of halo
losses in terms of large ratios of the maximum displacement to the
RMS size of the beam.

\appendix

\section{Coupled field equations}\label{coupled}

\noindent Starting from the Maxwell equations for the space-charge
field felt by our particle in the beam
\begin{eqnarray*}
\mathbf{\nabla}\cdot \mathbf{E}_{sc}(\mathbf{r},t)&=&
  \frac{\rho_{sc}(\mathbf{r},t)}{\epsilon_0}\\
\mathbf{\nabla}\times \mathbf{E}_{sc}(\mathbf{r},t)&=&
  - \,\partial_t \mathbf{B}_{sc}(\mathbf{r},t)\\
\mathbf{\nabla}\cdot \mathbf{B}_{sc}(\mathbf{r},t)&=&0\\
\mathbf{\nabla}\times \mathbf{B}_{sc}(\mathbf{r},t)&=&
  \mu_0 \mathbf{j}_{sc}(\mathbf{r},t)+\frac{\partial_t
  \mathbf{E}_{sc}(\mathbf{r},t)}{c^2}
\end{eqnarray*}
and introducing the corresponding electro--magnetic potentials
$(\mathbf{A}_{sc},\Phi_{sc})$ from
\begin{eqnarray*}
 \mathbf{B}_{sc}(\mathbf{r},t)&=&\nabla\times
 \mathbf{A}_{sc}(\mathbf{r},t)\\
 \mathbf{E}_{sc}(\mathbf{r},t)&=&-\partial_t\mathbf{A}_{sc}(\mathbf{r},t)
 -\nabla\Phi_{sc}(\mathbf{r},t)
\end{eqnarray*}
with the gauge condition
\begin{equation}\label{gauge}
  \nabla\cdot\mathbf{A}_{sc}(\mathbf{r},t)
+\frac{1}{c^2}\,\partial_t\Phi_{sc}(\mathbf{r},t)=0\,,
\end{equation}
through the usual procedure we get the wave equations
\begin{eqnarray}
 \nabla^2\mathbf{A}_{sc}(\mathbf{r},t)
-\frac{1}{c^2}\partial_t^2\mathbf{A}_{sc}(\mathbf{r},t)
   &=&-\mu_0\mathbf{j}_{sc}(\mathbf{r},t)\label{waveqA}\\
 \nabla^2\Phi_{sc}(\mathbf{r},t)
-\frac{1}{c^2}\partial_t^2\Phi_{sc}(\mathbf{r},t)
   &=&-\frac{\rho_{sc}(\mathbf{r},t)}{\epsilon_0}\label{waveqFi}
\end{eqnarray}
It is also well known (see for example \cite{landau}, chapter XV)
that the Schr\"odinger equation for spinless, charged particles in
an electro--magnetic field $(\mathbf{A},\Phi)$ is
\[
  i\alpha\partial_t\psi=
  \left[\frac{1}{2m}(i\alpha\nabla
-\frac{q_0}{c}\mathbf{A})^2+q_0\Phi\right]
  \psi\,.
\]
For a charged particle in the beam the electro--magnetic field is
the superposition of the space-charge potential
$(\mathbf{A}_{sc},\Phi_{sc})$ plus an external, control potential
$(\mathbf{A}_{ext},\Phi_{ext})$, and hence we obtain the
Schr\"odinger equation\refeq{schroedinger} that we rewrite here
for convenience:
\begin{eqnarray}\label{schroedappendix}
  i\alpha\partial_t\psi&=&\frac{1}{2m}
  \left[i\alpha\nabla
  -\frac{q_0}{c}(\mathbf{A}_{sc}+\mathbf{A}_{ext})\right]^2\psi\nonumber\\
  &&\nonumber\\
  &&\;\;\;\;\;\;\;\;+q_0(\Phi_{sc}+\Phi_{ext})\psi\,.
\end{eqnarray}
It is apparent now
that\refeq{gauge},\refeq{waveqA},\refeq{waveqFi}
and\refeq{schroedappendix} constitute a self--consistent system of
coupled, non linear differential equations for the fields $\psi$,
$\mathbf{A}_{sc}$ and $\Phi_{sc}$.

Since we are in a reference frame comoving with the beam, in the
following we will suppose to deal only with stationary wave
functions of the form\refeq{stationarywf}. In this case, as
already remarked, we get $\mathbf{j}_{sc}=0$, so that we can take
the trivial solution $\mathbf{A}_{sc}=0$ of\refeq{waveqA}. The
gauge condition\refeq{gauge} then implies that
$\partial_t\Phi_{sc}=0$ and the wave equation\refeq{waveqFi}
reduces itself to the Poisson equation for the electrostatic
potential. Finally, by supposing also $\mathbf{A}_{ext}=0$, our
system is reduced to only two coupled, non linear equations for
the couple $(u,\Phi_{sc})$, namely
\begin{eqnarray*}
Eu&=&- \frac{\alpha^2}{2m}\nabla^2u+q_0(\Phi_{ext}+\Phi_{sc})\,u\\
\nabla^2\Phi_{sc}&=&-
\frac{Nq_0}{\epsilon_0}\,\frac{|u|^2}{\|u\|^2}
\end{eqnarray*}
where we have defined
\[
\|u\|^2=\int_{\mathbf{R}^3}|u(\mathbf{r},t)|^2\,d^3\mathbf{r}\,.
\]
Then if we pass to the potential energies by putting
\[
V_{ext}(\mathbf{r},t)=q_0\Phi_{ext}(\mathbf{r},t)\,,\qquad
V_{sc}(\mathbf{r},t)=q_0\Phi_{sc}(\mathbf{r},t)
\]
our equations are reduced to
\begin{eqnarray}
&&\frac{\alpha^2}{2m}\nabla^2u+(E-V_{ext}-V_{sc})\,u=0\,,\label{selfconstat1}\\
&&\nabla^2V_{sc}= -
\frac{Nq_0^2}{\epsilon_0}\,\frac{|u|^2}{\|u\|^2}\,.\label{selfconstat2}
\end{eqnarray}
Remark that here the wave functions $\psi$ and $u$, while
certainly normalizable, are not supposed in general to be already
normalized. However the equations\refeq{selfconstat1}
and\refeq{selfconstat2}, albeit non linear, are apparently
invariant for the multiplication of $u$ by an arbitrary constant.
The equations\refeq{selfconstat1} and\refeq{selfconstat2} will be
used in the Section\myref{selfconsistent}.

\section{Cylindrical harmonic oscillator}\label{cylho}

\noindent In the model discussed in this paper we suppose that the
external potential $V_e$ is a cylindrically symmetric, harmonic
potential with a proper frequency $\omega$, so that the ground
state without space-charge interaction will have a variance
\[
\sigma^2=\frac{\alpha}{2m\omega}\,.
\]
This means that, in the usual cylindrical coordinates
$\{r,\varphi,z\}$ with $r=\sqrt{x^2+y^2}$, we have
\begin{equation}\label{cylindrHO}
V_e(r)=\frac{m}{2}\,\omega^2r^2=\frac{\alpha^2}{8m\sigma^4}\,r^2
\end{equation}
and the corresponding Schr\"odinger equation
\begin{eqnarray*}
i\alpha\partial_t\psi&=&-\frac{\alpha^2}{2m}\nabla^2\psi+V_e\psi\\
  &=&-\frac{\alpha^2}{2m}\left[\frac{1}{r}\,\partial_r(r\partial_r)
      +\frac{1}{r^2}\,\partial^2_{\varphi}+\partial^2_z\right]\psi+V_e\psi
\end{eqnarray*}
with periodic conditions at $z=\pm L/2$ (for a bunch of length
$L$) has the eigenvalues
\[
E_{nk}=(n+1)\,\alpha\omega+k^2\left(\frac{2\pi\sigma}{L}\right)^2\alpha
\omega
\]
and the normalized eigenfunctions
\[
\psi_{nkl}(r,\varphi,z,t)=
R_{nl}(r)\,\Phi_l(\varphi)\,Z_k(z)\,e^{-iE_{nk}t/\alpha}
\]
with
\begin{eqnarray*}
R_{nl}(r)&=&\sqrt{\frac{\left(\frac{n
-l}{2}\right)!}{\left(\frac{n+l}{2}\right)!}}\,
            \frac{e^{-r^2/4\sigma^2}}{\sigma}\left(\frac{r}{\sigma
\sqrt{2}}\right)^l
            \mathrm{L}^{(l)}_{\frac{n
-l}{2}}\left(\frac{r^2}{2\sigma^2}\right)\\
\Phi_l(\varphi)&=&\frac{e^{il\varphi}}{\sqrt{2\pi}}\\
Z_k(z)&=&\frac{e^{i2k\pi z/L}}{\sqrt{L}}
\end{eqnarray*}
where $L_p^{(q)}(x)$ are the generalized Laguerre polynomials and
\begin{eqnarray*}
n&=&0,1,2,\ldots\\
l&=&\left\{\begin{array}{ll}
               0,2,4,\ldots,n&\quad\mbox{if $n$ even}\\
               1,3,5,\ldots,n&\quad\mbox{if $n$ odd}
           \end{array}
    \right. \\
k&=&0,\pm1,\pm2,\ldots
\end{eqnarray*}
We suppose that, by neglecting the space-charge interaction and in
a comoving frame of reference, the system will be correctly
described by the ground state
\begin{equation}\label{appunpertground}
\psi_{000}(r)=\frac{e^{-r^2/4\sigma^2}}{\sigma\sqrt{2\pi L}}
\end{equation}
associated to the eigenvalue $E_{00}=\alpha\omega$.

\vspace{0.2cm}

\acknowledgments{We thank Armando Bazzani, Oliver
Boine-Frankenheim, Francesco Guerra, Ingo Hofmann, and Giorgio
Turchetti for many valuable discussions along the writing of this
paper. We also acknowledge the financial support from INFN --
Istituto Nazionale di Fisica Nucleare (experiment MQSA); from INFM
-- Istituto Nazionale per la Fisica della Materia; and from MIUR
-- Ministero dell'Istruzione, dell'Universit\`a e della Ricerca.}

\vfill\eject


\vfill

\begin{thebibliography}{99}

\bibitem{koziol} H. Koziol, Los Alamos M. P. Division Report No.
MP-3-75-1 (1975).

\bibitem{reiser} M. Reiser, C. Chang, D. Kehne, K. Low,
T. Shea, H. Rudd, and J.Haber, Phys. Rev. Lett.
{\bf 61}, 2933 (1988).

\bibitem{gluck94} R. L. Gluckstern, Phys. Rev. Lett. {\bf 73},
1247 (1994).

\bibitem{gluck95} R. L. Gluckstern, W.-H. Cheng and H. Ye,
Phys. Rev. Lett. {\bf 75}, 2835 (1995).

\bibitem{gluck96} R. L. Gluckstern, W.-H. Cheng, S. S. Kurennoy,
and H. Ye, Phys. Rev. E {\bf 54}, 6788 (1996).

\bibitem{okamoto97} H. Okamoto and M. Ikegami, Phys. Rev.
E {\bf 55}, 4694 (1997).

\bibitem{gluck98} R. L. Gluckstern, A. V. Fedotov,
S. S. Kurennoy, and R. Ryne, Phys. Rev.
E {\bf 58}, 4977 (1998).

\bibitem{wangler98} T. P. Wangler, K. R. Crandall, R. Ryne,
and T. S. Wang, Phys. Rev. ST-AB {\bf 1}, 084201 (1998).

\bibitem{gluck99} A. V. Fedotov, R. L. Gluckstern,
S. S. Kurennoy, and R.Ryne, Phys. Rev. ST-AB
{\bf 2}, 014201 (1999).

\bibitem{ikegami99} M. Ikegami, S. Machida, and T. Uesugi,
Phys. Rev. ST-AB {\bf 2}, 124201 (1999).

\bibitem{quiang00} J. Quiang and R. Ryne, Phys. Rev.
ST-AB {\bf 3}, 064201 (2000).

\bibitem{hofmann00} O. Boine-Frankenheim and I. Hofmann,
Phys. Rev. ST-AB {\bf 3}, 104202 (2000).

\bibitem{hofmann01} L. Bongini, A. Bazzani, G. Turchetti, and
I. Hofmann, Phys. Rev. ST-AB {\bf 4}, 114201 (2001).

\bibitem{hofmann02} A. V. Fedotov and I. Hofmann,
Phys. Rev. ST-AB {\bf 5}, 024202 (2002).

\bibitem{wangler} T. Wangler, {\it RF Linear Accelerators}
(Wiley, New York, 1998).

\bibitem{landau}
L. D. Landau and E. M. Lifshitz, {\it Physical Kinetics}
(Butterworth-Heinemann, Oxford, 1996).

\bibitem{ruggiero} F. Ruggiero, Ann. Phys. (N.Y.) {\bf 153},
122 (1984); F. Ruggiero, E. Picasso and L. A. Radicati, Ann.
Phys. (N. Y.) {\bf 197}, 396 (1990).

\bibitem{struckmeier} J. Struckmeier, Phys. Rev.
ST-AB {\bf 3}, 034202 (2000).

\bibitem{paul} W. Paul and J. Baschnagel, {\it Stochastic Processes:
From Physics to Finance} (Springer, Berlin, 2000)

\bibitem{cufaropre} N. Cufaro Petroni, S. De Martino, S. De Siena,
and F. Illuminati, Phys. Rev. E {\bf 63}, 016501 (2001); N.Cufaro
Petroni, S. De Martino, S. De Siena, and F. Illuminati, in {\it
Quantum aspects of beam physics 2K}, P. Chen ed. (World
Scientific, Singapore, 2002) p.507;

\bibitem{demartino}
S. De Martino, S. De Siena, and F. Illuminati, Physica A
{\bf 271}, 324 (1999).

\bibitem{nelson}
E. Nelson, {\it Dynamical Theories of Brownian Motion} (Princeton
University Press, Princeton N. J., 1967);
E. Nelson, {\it Quantum Fluctuations} (Princeton University Press,
Princeton N. J., 1985).

\bibitem{guerraphysrep}
F. Guerra, Phys.Rep. {\bf 77}, 263 (1981).

\bibitem{guerravariaz} F. Guerra and L. M. Morato, Phys. Rev. D
{\bf 27}, 1774 (1983).

\bibitem{albeverio} S. Albeverio, Ph. Blanchard and R.
H\o gh-Krohn, Expo. Math. {\bf 4}, 365 (1983).

\bibitem{fedele} R. Fedele, G. Miele and L.
Palumbo, Phys. Lett. A {\bf 194}, 113 (1994), and references
therein;
S. I. Tzenov, Phys. Lett. A {\bf 232}, 260 (1997).

\bibitem{pusterla} S. A. Khan and M. Pusterla, Eur. Phys. J. A
{\bf 7}, 583 (2000).

\bibitem{morato} L. Morato, J. Math. Phys. {\bf 23}, 1020 (1982).

\bibitem{madelung} E. Madelung, Z. Physik {\bf 40}, 332 (1926);
D. Bohm, Phys. Rev. {\bf 85}, 166 (1952); Phys. Rev. {\bf 85},
180 (1952).

\bibitem{cufarojpa} N. Cufaro Petroni, S. De Martino,
S. De Siena, and F. Illuminati, J. Phys. A {\bf 32},
7489 (1999).

\bibitem{qabp1} N. Cufaro Petroni, S. De Martino, S.
De Siena, and F. Illuminati, in {\it Quantum aspects of beam
physics}, P. Chen ed. (World Scientific, Singapore, 1999) p.710;
N. Cufaro Petroni, S. De Martino, S. De Siena, R. Fedele, F.
Illuminati, and S. I. Tzenov, in Proceedings of the {\it European
particle accelerator conference -- EPAC98}, S. Myers et al. eds.
(IoP Publishing, Bristol, 1998) p.1259.

\bibitem{bvth}
N. Cufaro Petroni and F. Guerra, Found. Phys. {\bf 25}, 297
(1995).

\bibitem{cufaropla} N. Cufaro Petroni, S. De Martino, and
S. De Siena, Phys. Lett. A {\bf 245}, 1 (1998).

\end{thebibliography}
\end{document}